\documentclass[twocolumn,aps,superscriptaddress,pre]{revtex4}

\usepackage{amsmath,amssymb,graphicx}
\usepackage{subfigure}
\usepackage{algorithmic}
\usepackage{enumerate}
\usepackage{times}
\usepackage{color}
\usepackage{soul}
\definecolor{yblue}{rgb}{0.06, 0.3, 0.57}
\definecolor{dgreen}{rgb}{0, 0.5, 0}
\usepackage[pdftex]{hyperref}
\hypersetup{colorlinks=true,linkcolor=blue,citecolor=blue,urlcolor=blue}

\graphicspath{{./fig-jpg/}}

\begin{document}

\title{Pairwise Interactions of Ring Dark Solitons with Vortices and
  other Rings: Stationary States, Stability Features and Nonlinear Dynamics}

\author{Wenlong Wang}
\email{wenlongcmp@scu.edu.cn}
\affiliation{College of Physics, Sichuan University, Chengdu 610065, China}

\author{Theodore Kolokolnikov}
\affiliation{Department of Mathematics and Statistics, Dalhousie University, Halifax, Canada}

\author{D. J. Frantzeskakis}
\affiliation{Department of Physics, National and Kapodistrian University of Athens,
Panepistimiopolis, Zografos, 15784 Athens, Greece}

\author{R. Carretero-Gonz{\'a}lez}
\affiliation{Nonlinear Dynamical Systems
Group,\footnote{\texttt{URL}: http://nlds.sdsu.edu}
Computational Sciences Research Center, and
Department of Mathematics and Statistics,
San Diego State University, San Diego, California 92182-7720, USA}

\author{P. G. Kevrekidis}
\email{kevrekid@math.umass.edu}
\affiliation{Department of Mathematics and Statistics, University of Massachusetts,
Amherst, Massachusetts 01003-4515 USA}

\begin{abstract}
In the present work, we explore analytically and numerically the
co-existence and interactions of ring dark solitons (RDSs) with other RDSs, 
as well as with vortices. The azimuthal instabilities of the rings are explored via
the so-called filament method. 
As a result of their nonlinear interaction, the vortices are found
to play a stabilizing role on the rings, yet their effect is not sufficient to
offer complete stabilization of RDSs. Nevertheless, complete
stabilization of the relevant configuration
can be achieved by the presence of external ring-shaped barrier potentials. Interactions 
of multiple rings are also explored, and their equilibrium positions (as a result of their own
curvature and their tail-tail interactions) are identified. In this
case too, stabilization is achieved via multi-ring external barrier potentials.
\end{abstract}

\maketitle

\section{Introduction}

Over the past two and a half decades,
Bose-Einstein condensation (BEC)~\cite{review_dalfovo,becbook1,becbook2} has 
served as a pristine example where numerous exciting features of nonlinear 
dynamics of coherent solitary wave structures can be studied and experimentally observed.
Solitary wave patterns, however, are not unique to BECs,
but rather appear generically in numerous other physical contexts including,
most notably, nonlinear optics~\cite{kivshar_agr}, 
plasma physics~\cite{Infeld} and water waves~\cite{abl3}. 
In the atomic physics setting,  while BEC
is known to be a fundamental phenomenon connected, e.g.,
to superfluidity and superconductivity~\cite{chap01:becgss},
BECs were only realized 70 years after their
prediction~\cite{chap01:anderson,chap01:davis,chap01:bradley}
and their broad relevance and impact were
rapidly acknowledged~\cite{chap01:rmpnlA,chap01:rmpnlB}.
Furthermore, the experimental observation of vortices~\cite{Matthews99,Madison00} and of  
highly ordered, triangular vortex lattices (predicted in Ref.~\cite{needcite}) in 
BECs~\cite{vle}, were found to be connected with the Kosterlitz-Thouless (KT) phase 
transition~\cite{needcite2,advanced}.
%
This transition has, indeed, found one of its most canonical
realizations in the context of BECs~\cite{dalikruger}.

In addition to being at the 
epicenter of some of the most important physical notions of the past two decades,
the coherent structures considered herein have also been recognized as
being of potential relevance to applications.
For instance, solitary waves have been argued to provide the potential
for
improved sensitivity in interferometers~\cite{negretti}
and precise force sensing applications.
Moreover, vortices may play a role similar to spinning black holes
in the so-called ``analogue gravity''.
This allows to observe, in
experimentally controllable environments, associated phenomena such as
the celebrated Hawking radiation~\cite{jeff}, or simpler ones such as super-radiant
amplification of sonic waves scattered from black holes~\cite{savage}.
It has also been recently argued that vortices of a
rotating BEC can collapse towards the generation of supermassive black holes~\cite{dasgupta}, 
and that supersonically expanding BECs can emulate properties of
an expanding universe in the lab~\cite{gretchen}.

Dark solitons have been of extensive interest within nonlinear
optics~\cite{Kivshar-LutherDavies} and over the last 20 years they
have also constituted a focal point in atomic BECs~\cite{djf}.
Vortices, on the other hand, have attracted considerable attention
broadly within nonlinear field theory~\cite{Pismen1999}, 
more specifically in nonlinear optics~\cite{YSKPiO} 
and especially in atomic BECs~\cite{fetter1,fetter2}.
These two structures share an important link: when dark
solitons are embedded in higher dimensions, they are prone to 
transverse (snaking) instability~\cite{kuzne,kidep}, and upon
undulations, eventually breakup, leading to spontaneous emergence of vortices. 
This feature has been experimentally demonstrated both in optics~\cite{yskchristou} 
and in BECs~\cite{watching}. 

Since the early work of Ref.~\cite{watching}, there have
been numerous proposals and associated experimental
efforts for creating vortices in BECs. These involved, among others, stirring the 
BEC~\cite{Madison00,Williams99}, dragging obstacles through it~\cite{kett99,BPA3},
using the phase imprinting method~\cite{S2Ket1}, and exploiting nonlinear 
interference~\cite{BPAPRL}.
The work of Ref.~\cite{BPA} enabled the use of the Kibble-Zurek (KZ) mechanism
to quench a gas of atoms rapidly across the BEC transition, and ``freeze''
phase gradients resulting in the formation of vortices. 
Subsequently, a technique was introduced that enabled the dynamical visualization of
vortices during an experiment~\cite{dshall}. This, in turn, led to a detailed
study of effective particle models that were used to better understand and predict 
various aspects of vortex dynamics observed in the 
experiments~\cite{dshall1,dshall2,dshall3,ex16_2,ex16_3b}.

It is interesting to explore the interaction of these two fundamental
entities that emerge in self-repulsive atomic condensates (and also,
similarly, in defocusing nonlinear optical media), namely the dark soliton
stripes and the vortices. Relevant studies, concerning the scattering 
of linear and nonlinear waves (i.e., dark solitons) by an optical vortex,  
were performed some time ago~\cite{nepom1,nepom2} in the context of nonlinear optics. 
In these works, the scattering process was identified as an optical analogue 
of the Aharonov-Bohm scattering, due to a classical analogy between the Aharonov-Bohm 
effect~\cite{abeffect} and the scattering of a linear wave by a vortex~\cite{nepom3}.
The works~\cite{nepom1,nepom2} focused on the scattering of a planar dark soliton 
stripe from a vortex and the resulting asymmetry of the scattered wave. While they 
observed the subsequent transverse instability, they did not try to quantify this 
type of phenomenology. Our aim here is to provide some controlled settings where 
the soliton-vortex interaction can be quantified. Specifically, we will consider the 
case of the ring dark soliton (RDS), predicted to occur in BECs in Ref.~\cite{rdsus}
(see also Refs.~\cite{kamchatnov,smirnov}), 
which has been an object of considerable study in its own right 
(see the reviews~\cite{djf,siambook} and references therein).
Various features of this structure have recently been explored,
including its modes of instability, via the so-called ``filament method''~\cite{aipaper},
its potential stabilization via external potentials~\cite{rcg:105} and
its extension to higher dimensions~\cite{babev}.

The main purpose of this work is to explore the interplay of the RDS with the vortex,
as well as with other RDSs. Part of the scope of this work is to
examine whether, e.g., the presence of the vortex can affect the stability of
the RDS and {accelerate} or {avert} (or delay) the transverse instability. 
This interaction is an especially interesting one because it is effectively 
one ``across dimensions'', as we are exploring the interplay of an effective point particle
(i.e., the vortex) with a quasi one-dimensional (1D) structure (the RDS). Indeed,
our theoretical analysis leads to the conclusion that the effect
of the vortex is in the direction of stabilizing the RDS and, 
the higher the vortex charge, the stronger this stabilization
effect. Nevertheless, this effect unfortunately weakens over increasing
chemical potential, and becomes negligible in the Thomas-Fermi (TF) large density 
(large chemical potential) limit. We also explore the case where additional RDSs may exist.
This is analogous to what was done, e.g., for dark solitons 
in quasi-1D settings~\cite{djf,siambook}, as well as for
multiple solitonic stripes, a case studied more recently in Ref.~\cite{george}.
Here, we find the equilibrium positions arising from the 
interplay of the RDSs with the confining potential and with each other. 
Finally, for both settings, we propose a method of stabilizing the structures 
of interest, relying on the use of external ring-shaped barrier potentials, 
which may enable their experimental realization. It is important to 
highlight here a report of experimental realization of such structures~\cite{bigelow}, 
that suggests they can be observed not only in a single- but also
in a multi-component BEC (in the form of dark-bright soliton states~\cite{siambook}). 

Our presentation is structured as follows. In Sec.~II we introduce the model and describe 
its basic features, while Sec.~III is devoted to our analytical results; these concern 
existence and stability results for a RDS with a central vortex, as well as multiple RDS 
states. In Sec.~IV we present our numerical results, while in Sec.~V, we summarize our 
conclusions and discuss possible relevant themes for future studies. 

\section{Model and Setup}

Our starting point, in both our analytical and numerical
considerations, is the following dimensionless Gross-Pitaevskii equation (GPE) 
in two dimensions:
\begin{eqnarray}
i \frac{\partial u}{\partial t} = -\frac{1}{2} \nabla^2 u+V u +|u|^2 u,
\label{mgpe}
\end{eqnarray}
where $u(x,y,t)$ is the macroscopic wavefunction, and $V=(1/2)\Omega^2r^2$ 
(with $r^2\equiv x^2+y^2$) is the 
radially symmetric harmonic potential; for the relevant adimensionalization, 
see, e.g., Ref.~\cite{siambook}. It is convenient to set $\Omega=1$ by scaling, 
without loss of generality. Using $u(x,y,t)=u_0(x,y) \exp(-i\mu t)$, with $\mu$ being 
the chemical potential, one obtains the time-independent GPE for $u_0(x,y)$:
\begin{eqnarray}
\mu u_0=-\frac{1}{2} \nabla^2 u_0 +V u_0 +|u_0 |^2 u_0.
\label{tigp}
\end{eqnarray}

In this work, we focus on multiple RDS structures, and also structures with a 
nesting central vortex of arbitrary charge. Interestingly, these type of structures 
have their respective linear limits as the linear two-dimensional (2D) quantum harmonic 
oscillator states in polar coordinates. Specifically, the normalized linear states 
take the following form (see, e.g., Ref~\cite{cct}):
\begin{equation}
\varphi_{n_r,n_{\theta}}(r,\theta) = \sqrt{\frac{n_r!}{\pi (n_r+|n_{\theta}|)!}} 
L_{n_r}^{|n_{\theta}|}(r^2) r^{|n_{\theta}|} e^{-\Omega r^2/2}e^{in_{\theta} \theta},
\label{linearstates}
\end{equation}
where $L$ is the associated Laguerre polynomial. The corresponding
eigenenergies are $E_{n_r,n_{\theta}}=(2n_r+|n_{\theta}|+1)\Omega$,
where $n_r, |n_{\theta}|=0, 1, 2, \ldots$. The two quantum numbers, in
turn, represent the number of circular nodes, and the angular 
$2\pi$-phase winding number or the topological charge.

The above states are particularly useful for our numerical procedure
identifying such states. More concretely, our numerical simulation includes 
identifying these stationary states, studying
their stability spectra via the so-called Bogolyubov-de Gennes (BdG)
analysis~\cite{siambook}, and, finally, identifying their dynamics. The states are
continued in chemical potential from the respective linear limits to
the large-density, TF regime. Because of the symmetry of the states up
to a topological charge, we compute the stationary states by solving the 
1D radial equation. The BdG spectrum is then computed using 
the so-called partial-wave method~\cite{DSS,PW2}. This consists of the decomposition 
of the form:
\begin{align}
  u(x,y,t)= & e^{-i\mu t} e^{i S \theta} \Big[ u_0(r) + \big( a_m(r) e^{i m
  \theta} e^{\lambda t} \nonumber \\
  & + b_m^*(r) e^{-i m \theta} e^{\lambda^* t} \big) \Big],
  \label{extra1}
\end{align}
and obtaining a radial eigenvalue problem for the eigenvector pair
$(a_m(r),b_m(r))$ in the case of each, integer, angular
(Fourier) mode, whose eigenvalues can be collected
to formulate the full spectrum of (generally complex) eigenvalues
$\lambda$. Here, $S$ plays the same role as the quantum number $n_{\theta}$ in
the linear case mentioned above, representing the topological charge.
In this work, we collect the low-lying modes $m=0, 1, 2, \ldots, 10$~\cite{aipaper}.
It is worth mentioning that this is much more efficient
than a direct 2D computation, allowing us to study stationary states
deep in the TF regime with improved  accuracy. When a full 2D state is
needed, e.g., for dynamics, it is initialized using the cubic spline interpolation.
Next, before focusing on these numerical results, let us discuss
reduction methods enabling us to understand the effective behavior
of rings in the presence of vortices, as well as in the presence of other rings. 

\section{Analytical Results}

\subsection{Angular instability of a RDS with a central vortex}

In the case of a ring harboring a vortex in its center, our
starting point will be  the Hamiltonian (energy) of the system, as in
the filament method of Ref.~\cite{aipaper}:
\begin{eqnarray}
H=\iint \left[ \frac{1}{2}\left\vert \nabla u\right\vert ^{2}+\frac{1}{2}\left(
\mu-V(r)-\left\vert u\right\vert ^{2}\right)^{2} \right] dxdy.
  \label{extra2}
  \end{eqnarray}
We use the ansatz
\begin{eqnarray}
u&=&e^{-i\mu t}e^{iS\theta}w, \nonumber \\ 
w&\equiv& 
iR_{t}+A\tanh\left(A(r-R)\right), 
  \label{extra3}
  \end{eqnarray}
where $R=R(t,\theta)$ is the ring radius, $R_t \equiv \partial R/\partial t$ is the radial 
velocity of the ring, and $A$ is the ring's depth, given by:
\[
A=\left(  \mu-V(R)-R_{t}^{2}\right)^{1/2}. 
\]
The above ansatz combines the 
radial analogue of the dark soliton, represented by the function $w$, and a vortex,   
represented by a simple phase profile $\exp(i S \theta)$ and neglecting the associated 
density dip ---as is often done in similar methodologies~\cite{fetter1,fetter2}. 

Performing the relevant integrals, we obtain that, up to a constant $C$, 
the Hamiltonian becomes:
%
\begin{eqnarray}
H=\int\left(  \frac{4}{3}RA^{3}\left(1+\frac{R_{\theta}^{2}}{2R^{2}}\right)
-\frac{S^{2}A}{R}\right)  d\theta+C.
  \label{extra4}
\end{eqnarray}
The dynamics of the ring is determined by the conservation of energy, i.e., setting $H_{t}=0.$
Through lengthy, but straightforward calculations, similar to those presented
in Ref.~\cite{aipaper}, 
we obtain the following equation for the ring radius $R$:
%
%
\begin{align}
&R_{tt}\left(  -3RA+\frac{3}{4}\frac{S^{2}}{AR}\right)  +R_{\theta\theta
}\left(  -\frac{A^{3}}{R}\right)  +A^{3} \nonumber \\
&-RA\frac{3}{2}V^{\prime}(R)+\frac
{3}{4}S^{2}\left(  \frac{V^{\prime}(R)}{2AR}+\frac{A}{R^{2}}\right) =0.
\label{1100}%
\end{align}
Considering the case of a parabolic trap, one has:
\[
V(R)=\frac{1}{2}\Omega^{2} R^{2},\ \ A^{2}\sim\mu-V(R),
\]
and, hence, Eq.~(\ref{1100}) can be rewritten  as:
\begin{eqnarray}
R_{\theta\theta}+f(R)R_{tt}+g(R)=0,
  \label{extra5}
  \end{eqnarray}
where:
\begin{align*}
f(R)  & =3\frac{R^{2}}{A^{2}}-\frac{3}{4}\frac{S^{2}}{A^{4}}, 
\nonumber
\\
\nonumber
\frac{g(R)}{R}  & =-1+R^{2}A^{-2}\frac{3}{2}\Omega^{2}-\frac{3}{4}%
S^{2}\left(  \frac{\Omega^{2}}{2A^{4}}+\frac{A^{-2}}{R^{2}}\right).
\end{align*}

Let $R_{0}$ be the root of $g$, i.e., $g(R_{0})=0$.
%
%
Then, using the ansatz:
\[
R=R_{0}+\varepsilon e^{in\theta}e^{i\omega t},
\]
and substituting in Eq.~(\ref{extra5}), we linearize (for $\varepsilon$ small) 
around $R_{0}$
and obtain:
\[
\omega^{2}=-\frac{n^{2}}{f(R_{0})}+\frac{g^{\prime}(R_{0})}{f(R_{0})}.
\]
It is worthwhile to note that in the absence of the topological charge 
(i.e., for $S=0$), the above equation reduces to the one 
presented in Ref.~\cite{aipaper}, yet here the expression is systematically
generalized for arbitrary vortex charges $S$.

Lastly, for the present considerations, it is possible to expand the
above expressions in the TF limit of large $\mu$. Series expansions up to two orders then
yield:
\begin{eqnarray}
R_{0}^{2}&\sim&\frac{\mu}{2\Omega^{2}}+\frac{1}{\mu}S^{2}+O(\mu^{-3}),
\label{neweq1}
\\ 
\nonumber
g^{\prime}(R_{0})&\sim&\frac{8}{3}+\frac{32}{8}\frac{\Omega^{2}}{\mu^{2}}S^{2},%
\label{neweq2}
\\ 
\label{neweq3}
f(R_{0})&\sim&\frac{2}{\Omega^{2}}+\frac{4S^{2}}{\mu^{2}},%
\end{eqnarray}
and thus finally lead to:
\begin{equation}
\omega^{2}\sim\frac{\Omega^{2}n^{2}}{2}\left(- 1+2\frac{\Omega^{2}S^{2}}%
{\mu^{2}}\right)  +\frac{4}{3}\Omega^{2}\left(  1+2\frac{S^{2}\Omega^{2}}%
{\mu^{2}}\right).
\label{prediction}
\end{equation}
It is worthwhile to note that {\it both} contributions of $S$ arise
with a positive sign, increasing the value of the squared eigenfrequency 
$\omega^2$. Recalling that the latter is 
related to the
corresponding squared eigenvalue according to $\omega^2=-\lambda^2$, 
one may conclude that the presence of the topological charge
{\it alleviates} the instability of the RDS. Nevertheless,
the relevant effect weakens rapidly with increasing chemical potential,
and disappears deeply inside the TF limit with $\mu \rightarrow \infty$. 
Our observations below support this feature, 
showcasing however 
that it is not possible for this partial stabilization to ``suffice'' towards
countering the azimuthal destabilization effects.

\subsection{Multiple RDSs}

We start with the stationary GPE~(\ref{tigp}) for a purely radial profile $u_0(r)$, 
which, when expressed in polar coordinates, yields:
\begin{eqnarray}
&\frac{1}{2}\left(u_{0,rr}+\frac{1}{r}u_{0,r}\right)+f(r)u_0-u_0^{3}=0,
\label{gper} \\ 
&f(r)=\mu-\frac{1}{2}\Omega^{2} r^{2}.
\label{fofr}
\end{eqnarray}
We consider a single stationary RDS of 
\emph{large} radius $r=r_{0}$. 
Then, expanding near $r_{0}$, i.e., $r=r_{0}+y$, we write
\[
u_0=U_{0}(y)+P(y),
\]
where $P(y)\sim O(1/r_{0})$ contains higher-order corrections. 
Then, at the leading order, Eq.~(\ref{gper}) yields the equation:
\begin{equation}
\frac{1}{2}U_{0yy}+f_{0}U_{0}-U_{0}^{3}=0,
\label{ode1}
\end{equation}
where $f_{0}=f(r_{0})$. Hence, at this order, we obtain via Eq.~(\ref{ode1}),  
the exact stationary RDS solution:  
\begin{equation}
U_{0}(y)=\sqrt{f_{0}}\tanh\left(\sqrt{f_{0}}y\right).
\label{strds}
\end{equation}
The next order equation reads:
\begin{equation}
\frac{1}{2}P_{yy}+f(r_{0})P-3U_{0}^{2}P+\frac{1}{2r_{0}}U_{0y}+f_{0}^{\prime}yU_{0}=0, 
\label{1058}%
\end{equation}
where $f_{0}^{\prime}=f^{\prime}(r_{0})$. Employing the Fredholm alternative,  
one may obtain the solvability condition of Eq.~(\ref{1058}), upon  multiplying it  
by $U_{0y}$ and integrating over $y$. This yields:
\begin{equation}
\int_{-\infty}^{\infty} \left( \frac{1}{2r_{0}}U_{0y}^{2}
+f_{0}^{\prime}yU_{0}U_{0y} \right) dy=0.
\label{1037}%
\end{equation}
Then, we compute the integrals, namely:
\begin{equation}
\int_{-\infty}^{\infty}U_{0y}^{2} dy=\frac{4}{3}f_{0}^{3/2},\quad \int
_{-\infty}^{\infty}yU_{0}U_{0y} dy=f_{0}^{1/2}, \label{1038}
\end{equation}
and substitute the relevant results into Eq.~(\ref{1037}). Furthermore, employing Eq.~(\ref{fofr}), 
and after some algebra, we obtain the radius of the stationary RDS, namely:
\begin{equation}
r_{0}=\sqrt{\frac{\mu}{2\Omega^{2}}}=\frac{R_{\rm TF}}{2}, 
\label{r0}%
\end{equation}
where we have introduced the TF radius $R_{\rm TF}=\sqrt{2\mu}/\Omega$ 
measuring the spatial extent of the background solution where the RDS is embedded.
This result recovers the one obtained previously in Refs.~\cite{aipaper,rcg:105,gth}.

\subsubsection{The two-RDS state}

We proceed with the stationary two-RDS state, which may be described by an
ansatz of the form:%
%
%
\begin{align}
u(r) = & \sqrt{f(r)}\Big[  \tanh\big(  \sqrt{f(r)} (  r-r_{+} ) \big) 
\nonumber
\\  
\nonumber
&-\tanh \big(  \sqrt{f(r)}(  r-r_{-} )  \big)  +1\Big],
\end{align}
where $r_{\pm}$ are the rings' interface locations, to be determined. 
One expects that, to the leading order, the rings' locations will be close to 
$r_{0}$ found in Eq.~(\ref{r0}). A more precise calculation of their relative positions 
requires the examination of their mutual interaction. Similarly to their purely 1D 
counterparts~\cite{ourmarkus3}, the two RDSs interact through their 
exponentially small tails, which determines their position relative to $r_{0}$.

\begin{figure*}
\includegraphics[width=\textwidth]{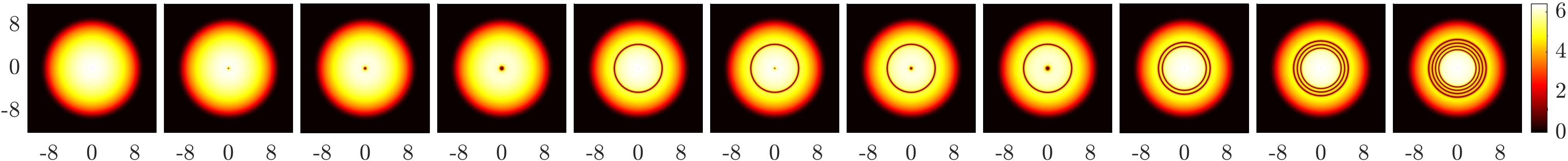}
\includegraphics[width=\textwidth]{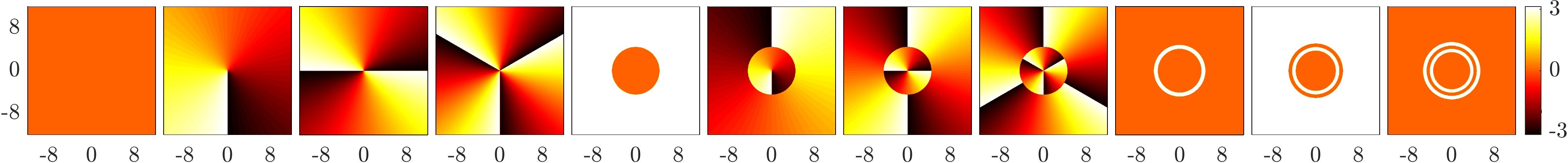}
\caption{(Color online) 
Typical stationary state profiles for the ground state, and a number of prototypical solitary wave states with various numbers of circular (ring) nodes, and a central vortex of various charges. The top panels depict their magnitudes, while the bottom panels depict their phases, the states here are numerically exact solutions at the chemical potential $\mu=40$. From left to right, we have the ground state, vortex states of charges $1, 2, 3$, the RDS 
state, one RDS with a central vortex of charges $1, 2, 3$, two, three, and four RDSs, respectively. Multiple RDSs can also nest a central vortex in the center, yet they are omitted here for clarity.
}
\label{states}
\end{figure*}

As before, we expand near each interface. First, expanding near $r_{+}$, and letting 
$r=r_{+}+y$, we write:
\begin{align*}
u    \sim & \sqrt{f_{+}}\tanh\left(  \sqrt{f_{+}}y\right)  
\nonumber
\\
\nonumber
&+\sqrt{f_{+}}\left[
1-\tanh\left(  \sqrt{f_{+}}\left(  y+r_{+}-r_{-}\right)  \right)  \right] = U_{0}+P 
\end{align*}
Let $L$ be any arbitrary constant, with $0\ll L\ll r_{+}-r_{-}$, 
such that $r_{+}-L\in\left(r_{-},r_{+}\right)$. Then, multiplying Eq.~(\ref{1058}) 
by $U_{0y}$ and integrating on $y\in\left(  -L,\infty\right)$, we obtain:
\[
\frac{1}{2}\left.  \left(  P_{y}U_{0y}-PU_{0yy}\right)  \right
\vert_{-L}^{\infty}+F(r_{+})=0, 
\]
where
\begin{eqnarray}
F(r) &=& \int_{-\infty}^{\infty} \left( \frac{1}{2r}U_{0y}^{2}+f^{\prime}(r)yU_{0}%
U_{0y} \right) dy  
\nonumber 
\\
\nonumber 
&=& f^{1/2}(r)\left(\frac{2}{3r}f(r)+f^{\prime}(r)\right), 
\end{eqnarray}
%
%
and $U_0$ is given by Eq.~(\ref{strds}). 
Next we evaluate the boundary term. Recalling that:
\[
\tanh z\sim\left\{
\begin{array}
[c]{c}%
1-2e^{-2z},\ \ \ ~z\rightarrow +\infty\\
-1+2e^{2z},\ \ \ \ \ z\rightarrow-\infty
\end{array}
\right.,   
\]
we find that, for $y$ near $-L$, one has:
\begin{align*}
U_{0}  &  \sim-\sqrt{f_{+}}+2\sqrt{f}\exp\left(  2\sqrt{f}y\right), 
\ y\ll-1,
\nonumber
\\
\nonumber
P  &  \sim2\sqrt{f}\exp\left[ -( y+r_{+}-r_{-})  2\sqrt {f}\right],\ y\ll-1.
\end{align*}
Hence, for $y\ll-1$, the boundary term reads:
\[
P_{y}U_{0y}-PU_{0yy}\sim-8f^{2}\exp\left[-2\sqrt{f}\left(r_{+}-r_{-}\right)\right].
\]
On the other hand, a similar expansion for positive $y$ yields $P_{y}%
U_{0y}-PU_{0yy}\sim0$ as $y\rightarrow\infty.$ In short, we obtain
\begin{eqnarray}
\frac{1}{2}\left.  \left(  P_{y}U_{0y}-PU_{0yy}\right) \right\vert_{-L}^{\infty}
&\sim& 16f^{2}\exp\left[  -2\sqrt{f}\left(  r_{+}-r_{-}\right)\right]
\nonumber \\
&\sim& -F(r_{+}) \label{extra10}
\end{eqnarray}
%
%
A similar computation near $r_{-}$ yields,
\begin{equation}
16f^{2}\exp\left(  -2\sqrt{f_{-}}\left(  r_{+}-r_{-}\right)  \right)
\sim+F(r_{-}).
\label{extra11}
\end{equation}
Assume that $f_{\pm}$ is large. Then, to leading order, this implies that
$r_{\pm}\sim r_{0}$ is the root of $F(r_{0})=0$ given by Eq.~(\ref{r0}). To
estimate its correction, we expand
\[
r_{\pm}=r_{0}+x_{\pm},\ \ \ \text{with }\left\vert x_{\pm}\right\vert \ll
r_{0}.
\]
We then estimate $f_{\pm}\sim f\left(  r_{0}\right)$ and $F(r_{+})\sim x_{\pm
}F^{\prime}(r_{0})$, so that Eqs.~(\ref{extra10}) and (\ref{extra11}) become:
\begin{eqnarray}
16f^{2}(r_{0})\exp(  -2\sqrt{f(r_{0})}(x_{+}-x_{-}  ))
\label{1054} 
&\!\sim\!&-F^{\prime}(r_{0})x_{+}, 
~~~~~
\\[1.0ex]
\label{1055} 
16f^{2}(r_{0})\exp(  -2\sqrt{f(r_{0})}(x_{+}-x_{-}  ))
&\!\sim\!&+F^{\prime}(r_{0})x_{-}. 
~~~~~
\end{eqnarray} 
Furthermore, we find:
\[
f\left(  r_{0}\right)  =\frac{3}{4}\mu,\ \ \ \ F^{\prime}(r_{0})=-\frac
{4}{\sqrt{3}}\mu^{1/2}\Omega^{2},
\]
so that Eqs.~(\ref{1054}) and (\ref{1055})\ can be written as:
\[
\exp\left(-bl\right)=al, \quad r_{\pm}=r_{0}\pm l/2 
\]
where
\begin{align}
a=3^{-5/2}4\mu^{-3/2}\Omega^{2}, \quad b=\sqrt{3\mu}. 
\label{ab}%
\end{align}
The relevant solution can also be written in terms of the Lambert
$w(\eta)$ function, which is defined as the inverse of $\eta(w)=w e^w$. 
More specifically, $l=(1/b) w(b/a)$. This is
reminiscent of the corresponding quasi-1D expression for the
equilibrium of two dark solitons in a trap~\cite{ourmarkus3}. Indeed, the intuitive
aspects of this formula reveal the role of the curvature toward
balancing the trap effect within $r_0$ (as for a single RDS)
and the role of inter-soliton interaction, in turn, balancing
the effect of the trap via the contribution within $l$. 

\begin{figure*}
\includegraphics[width=0.24\textwidth]{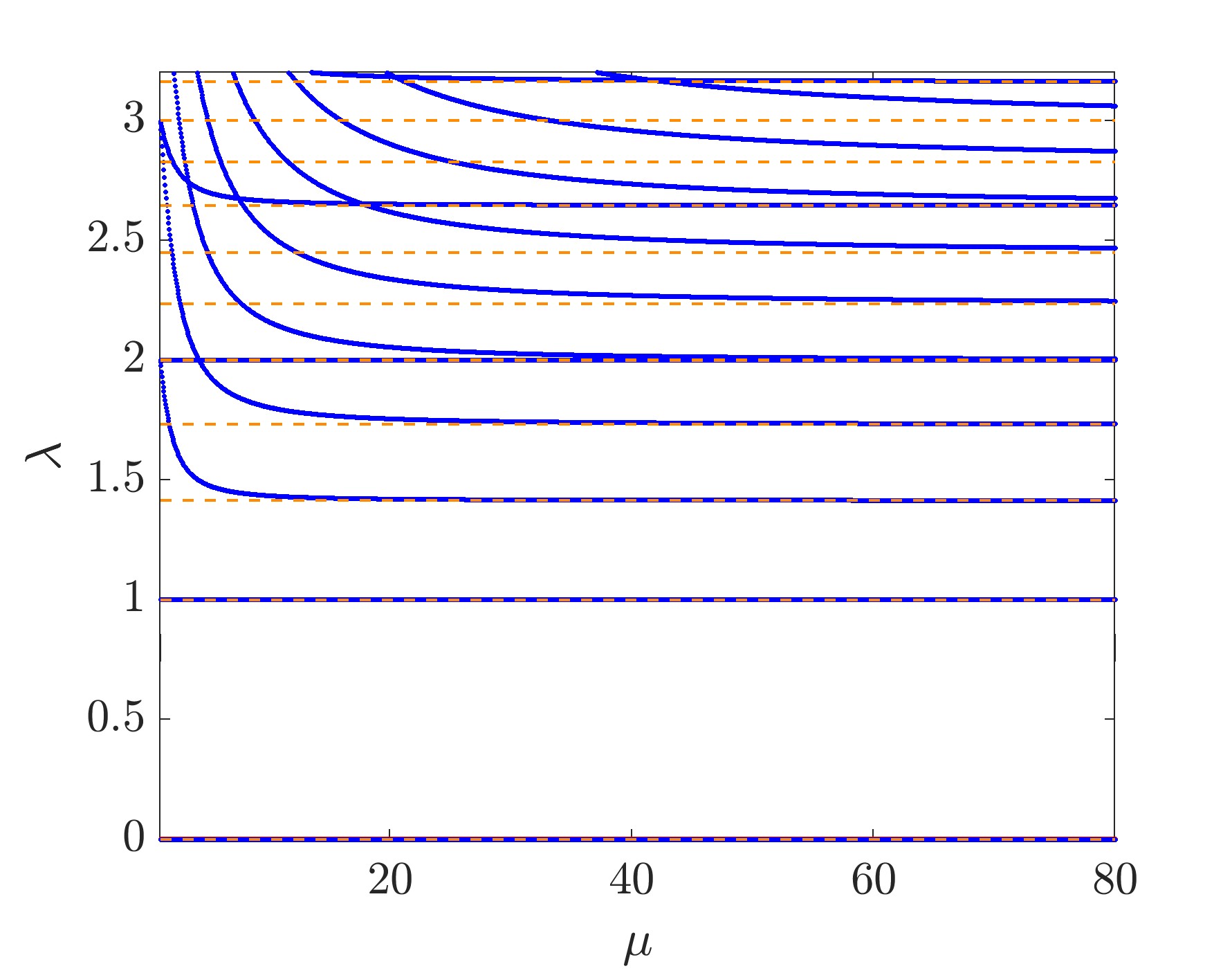}
\includegraphics[width=0.24\textwidth]{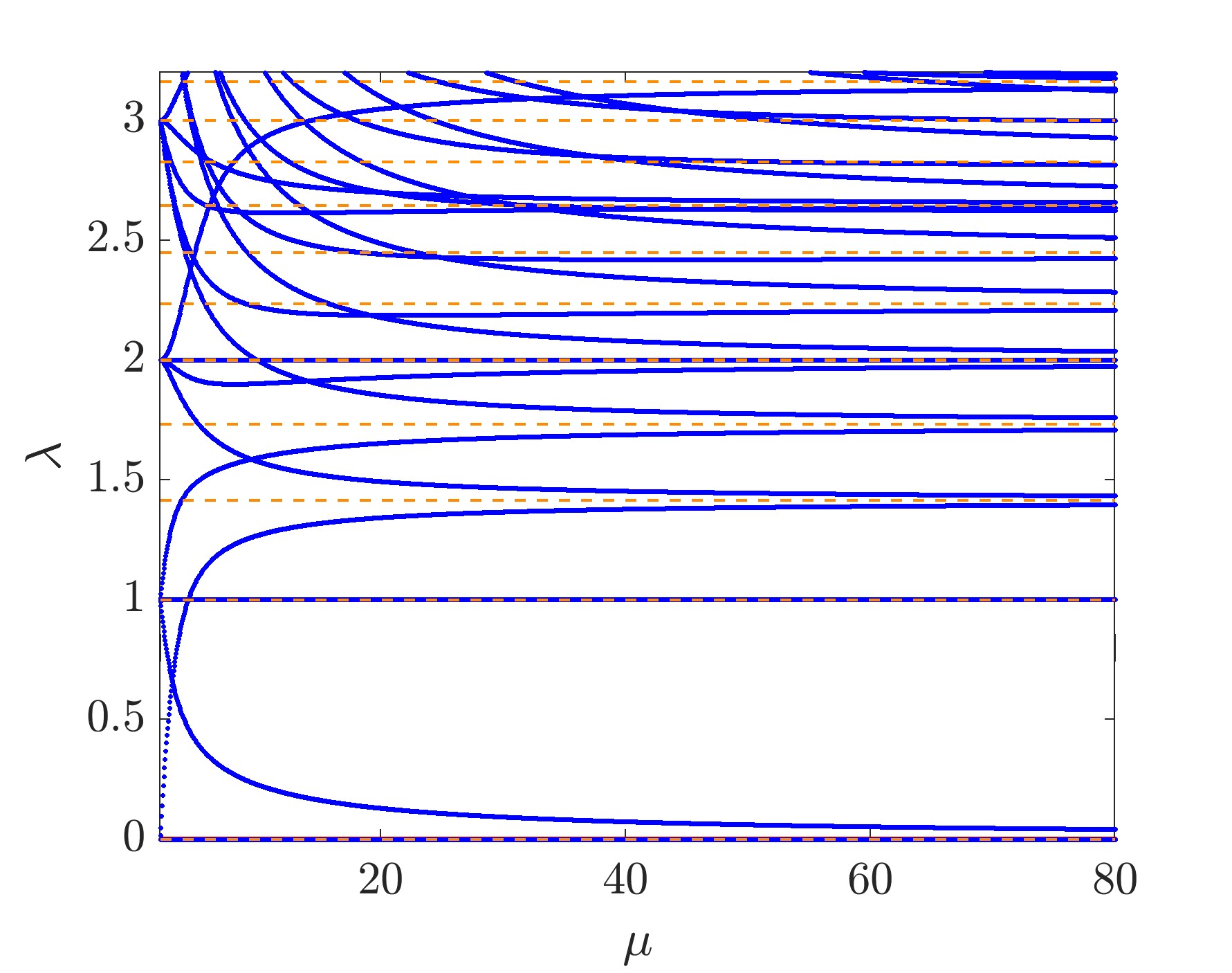}
\includegraphics[width=0.24\textwidth]{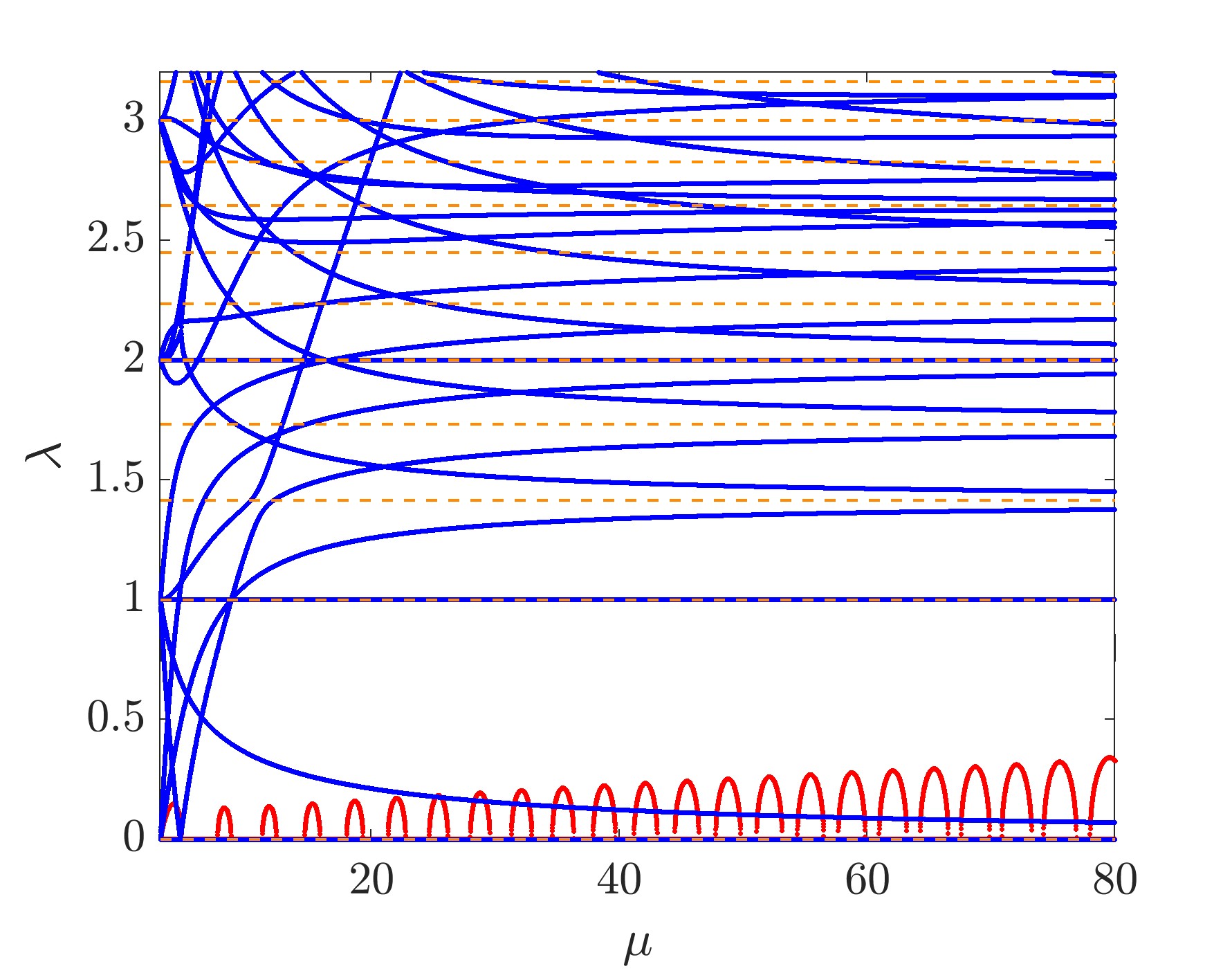}
\includegraphics[width=0.24\textwidth]{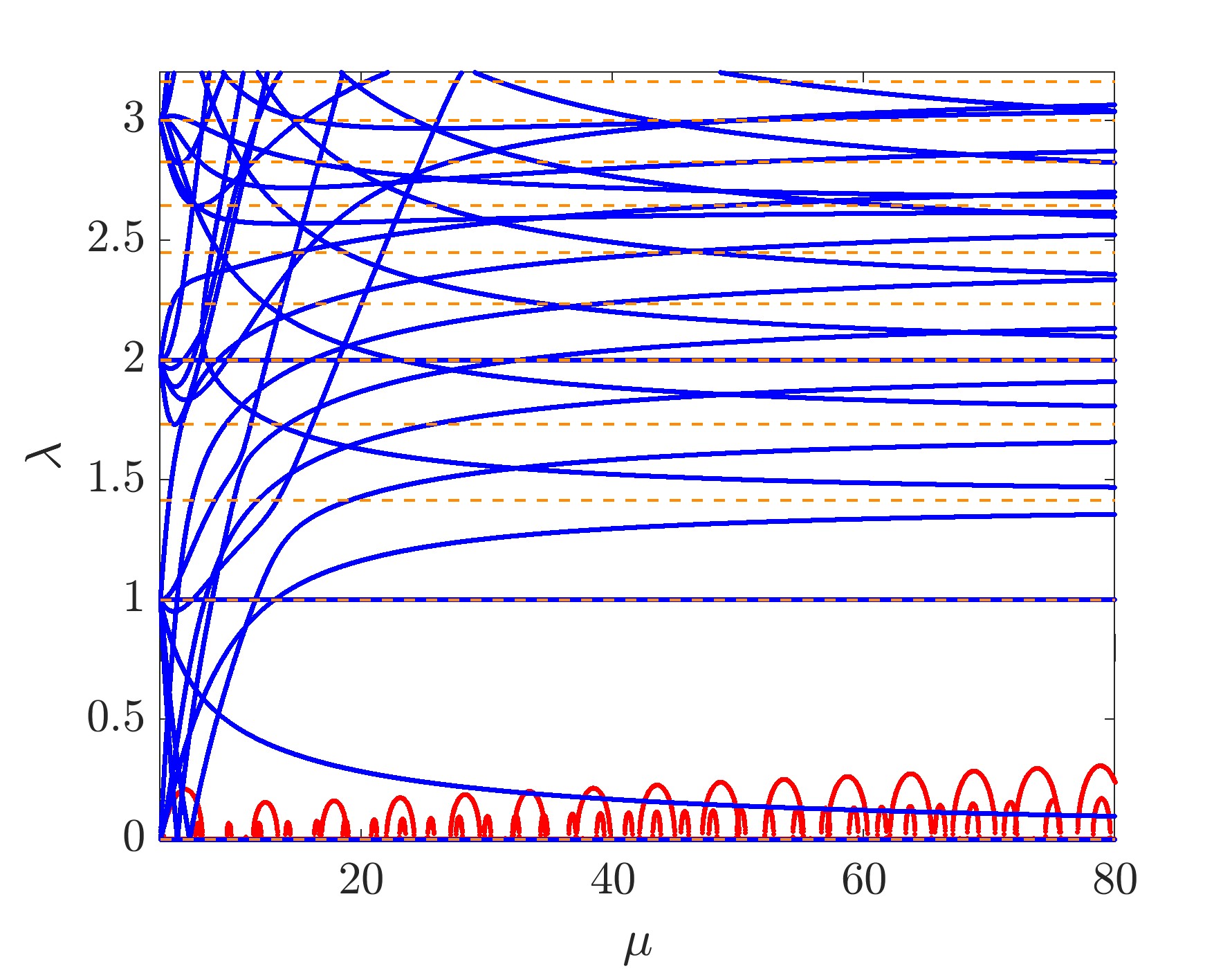}
\includegraphics[width=0.24\textwidth]{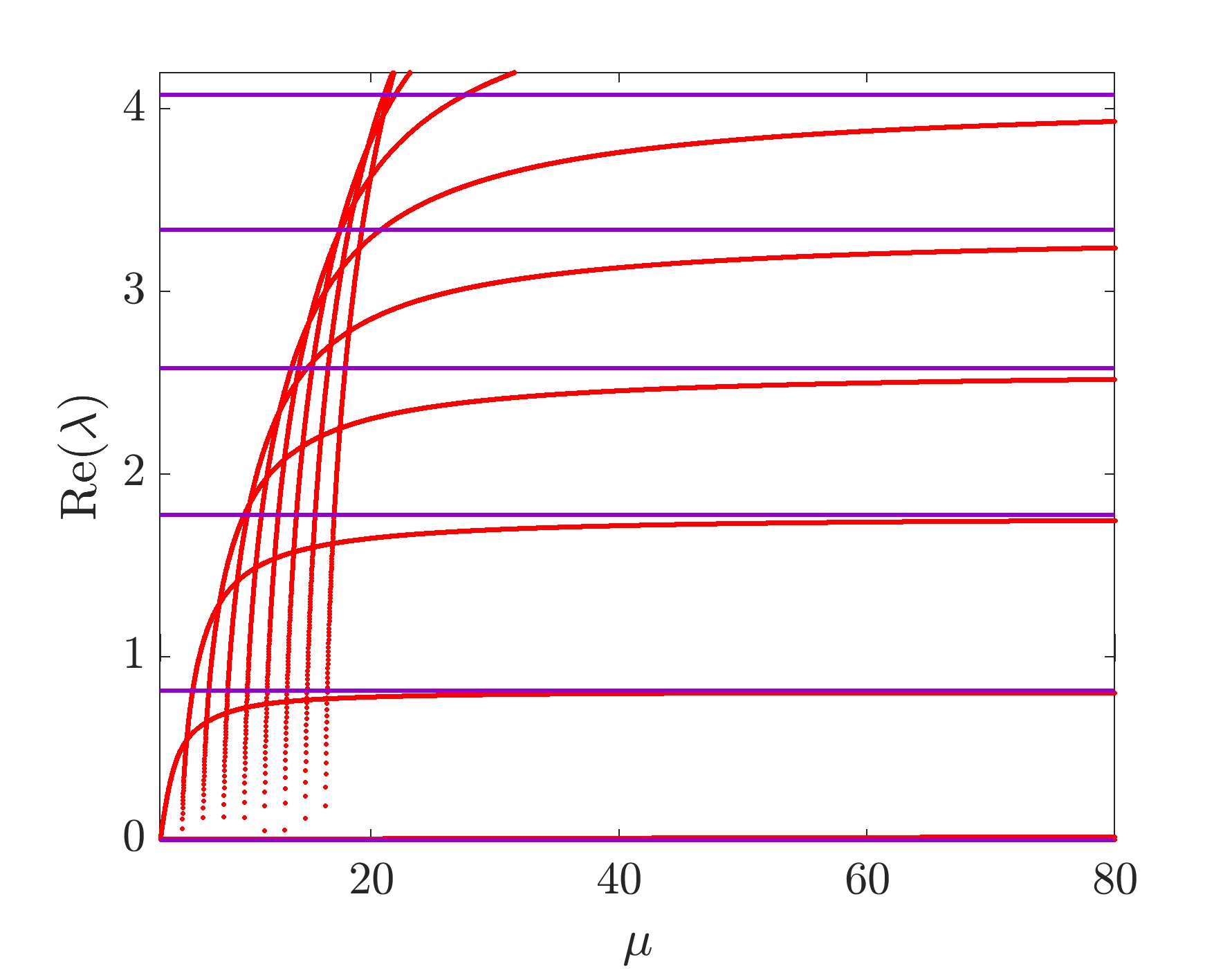}
\includegraphics[width=0.24\textwidth]{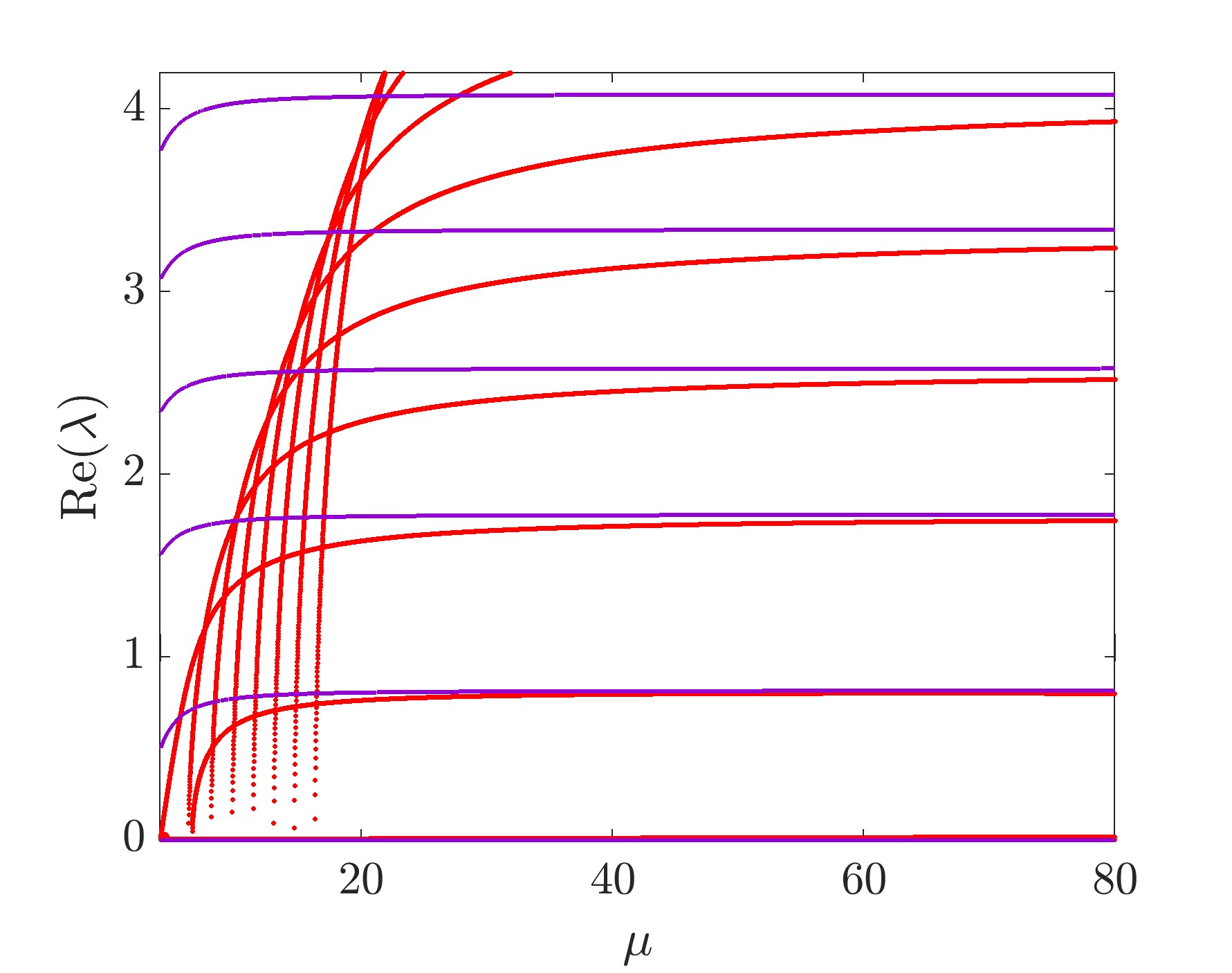}
\includegraphics[width=0.24\textwidth]{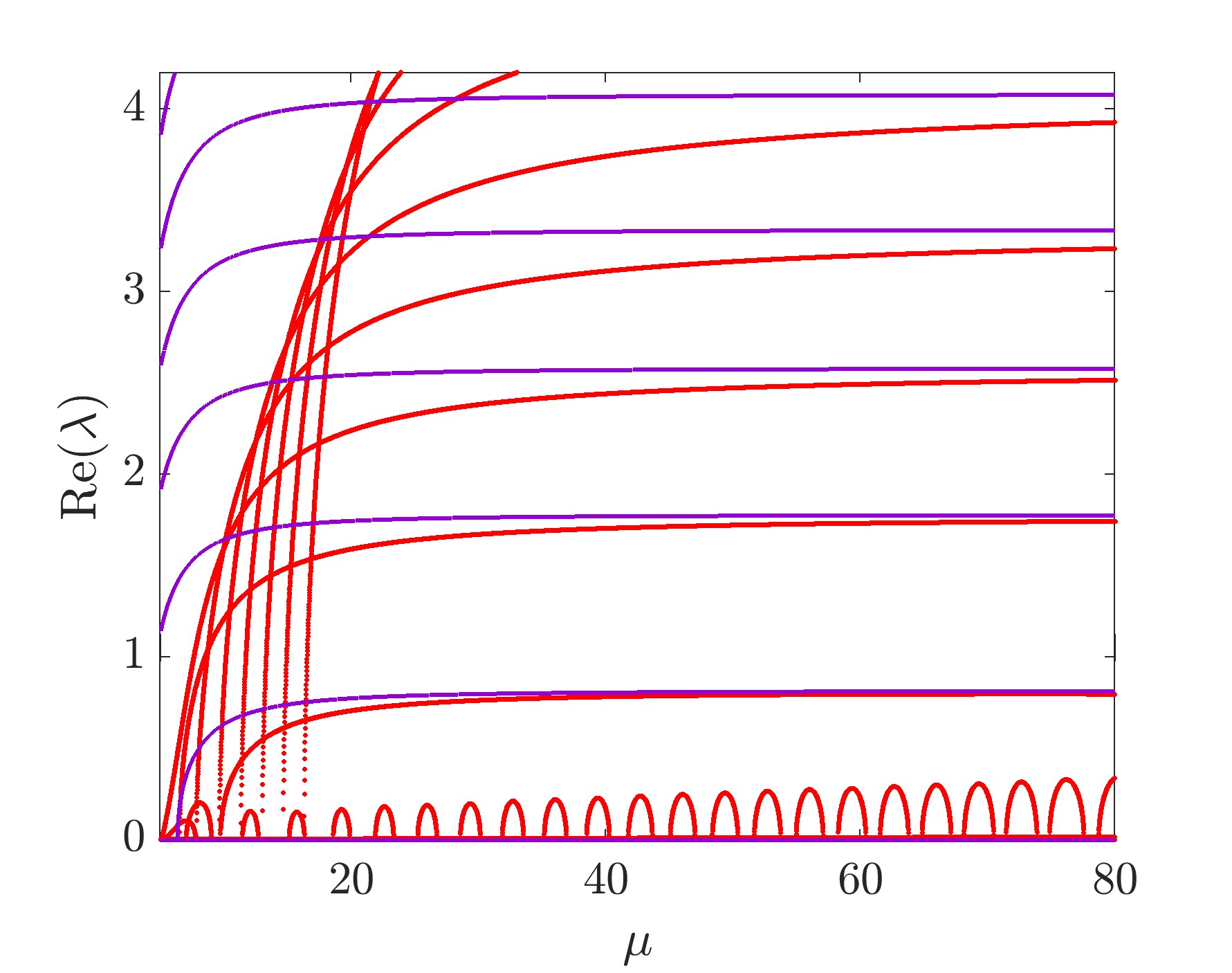}
\includegraphics[width=0.24\textwidth]{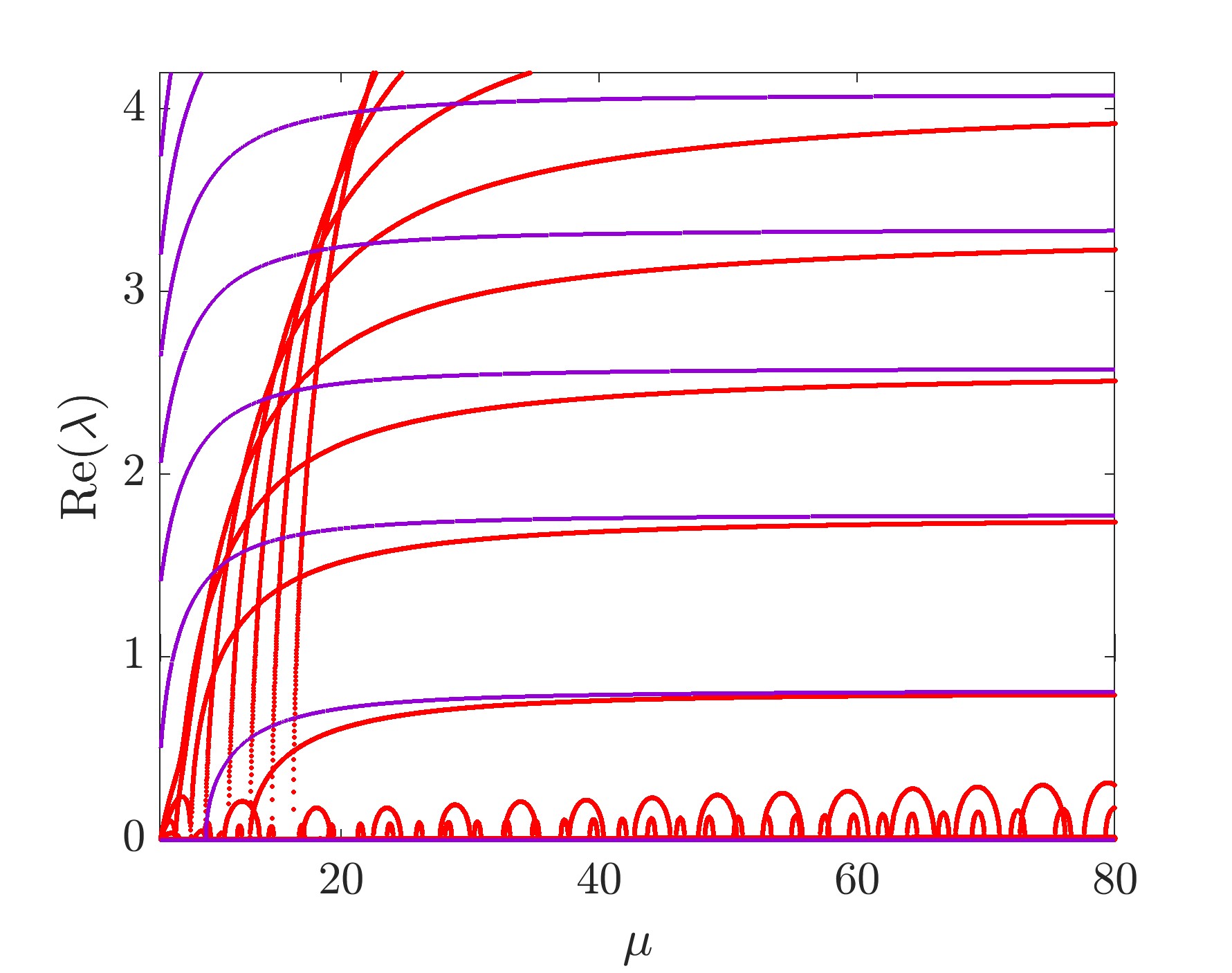}
\includegraphics[width=0.24\textwidth]{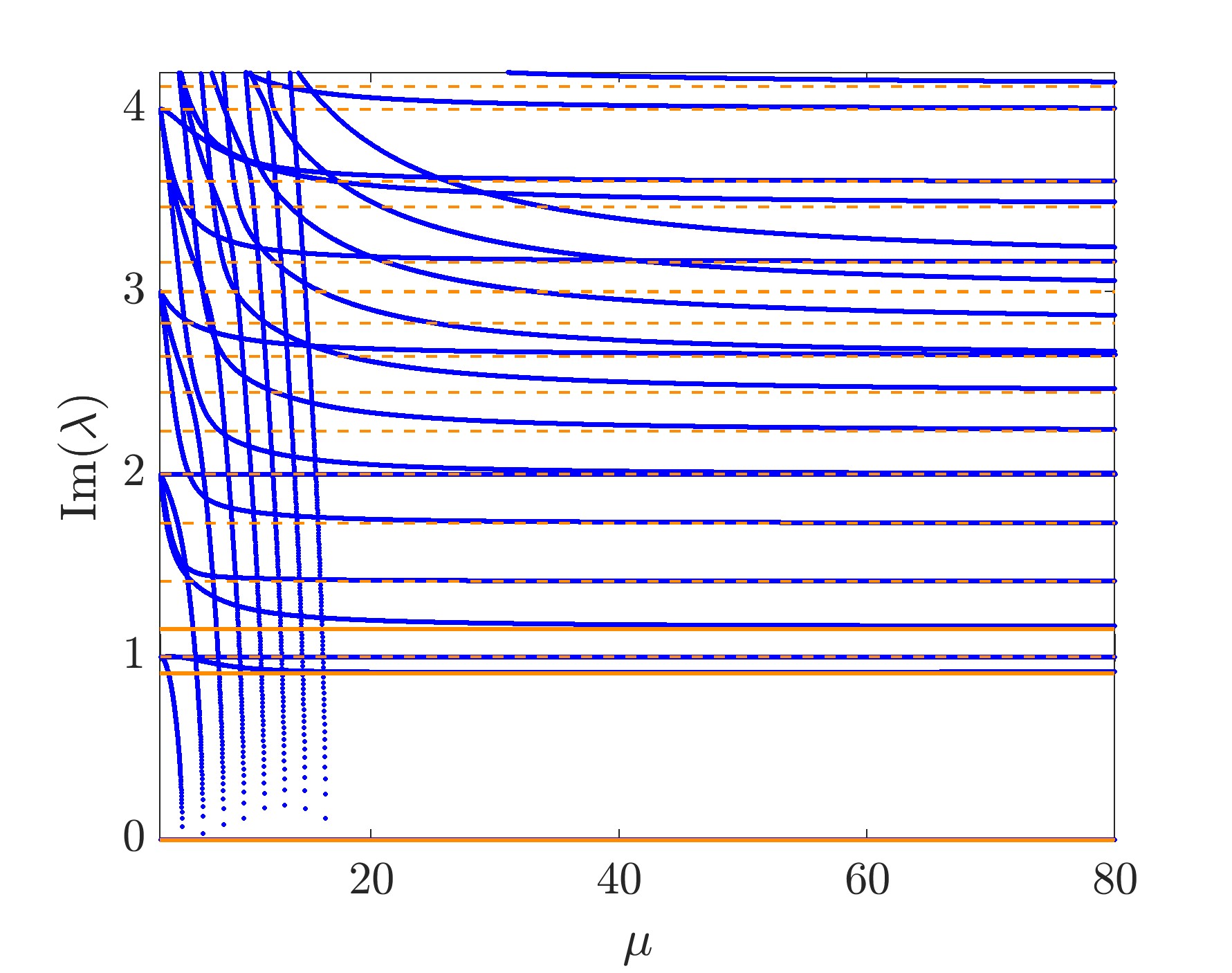}
\includegraphics[width=0.24\textwidth]{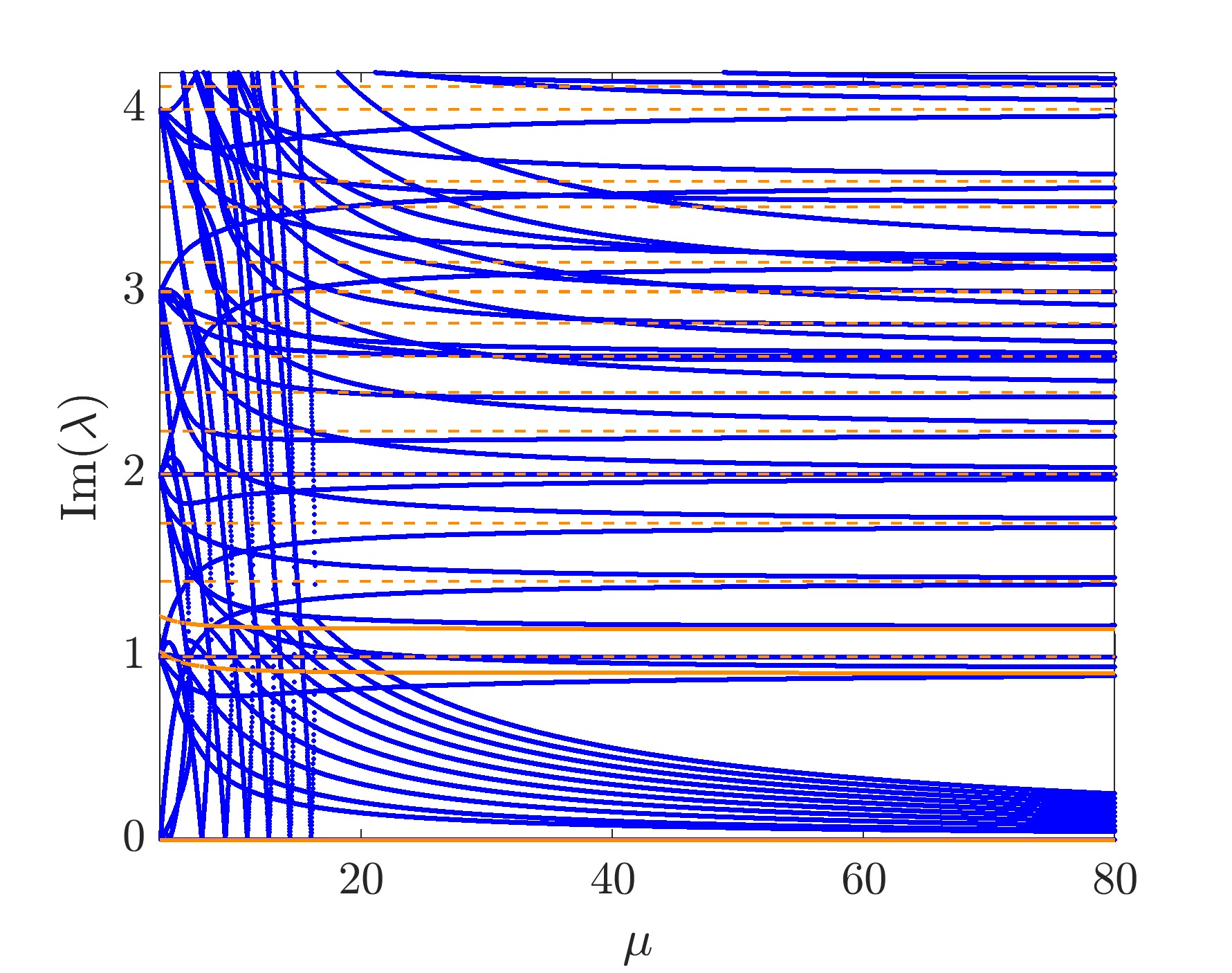}
\includegraphics[width=0.24\textwidth]{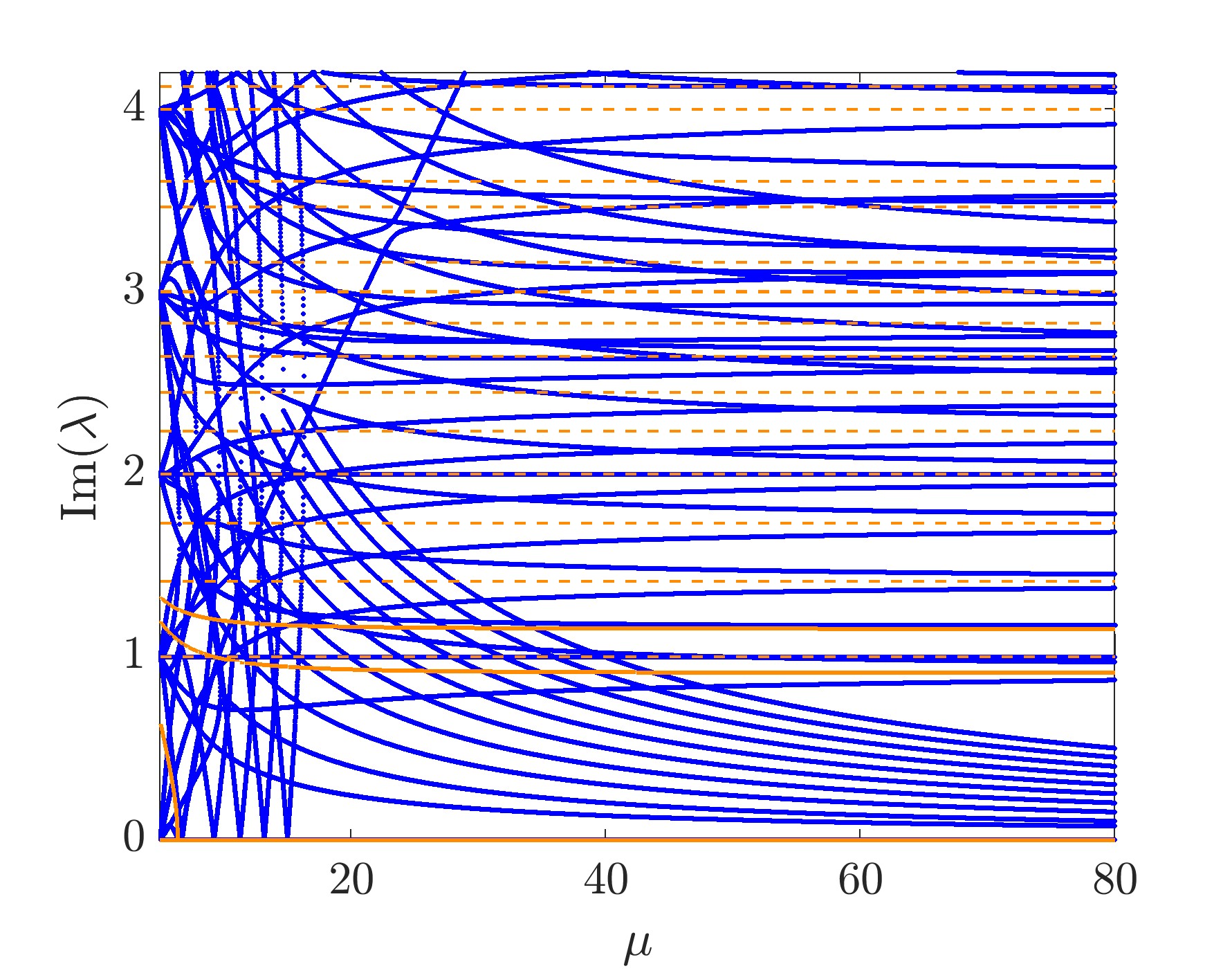}
\includegraphics[width=0.24\textwidth]{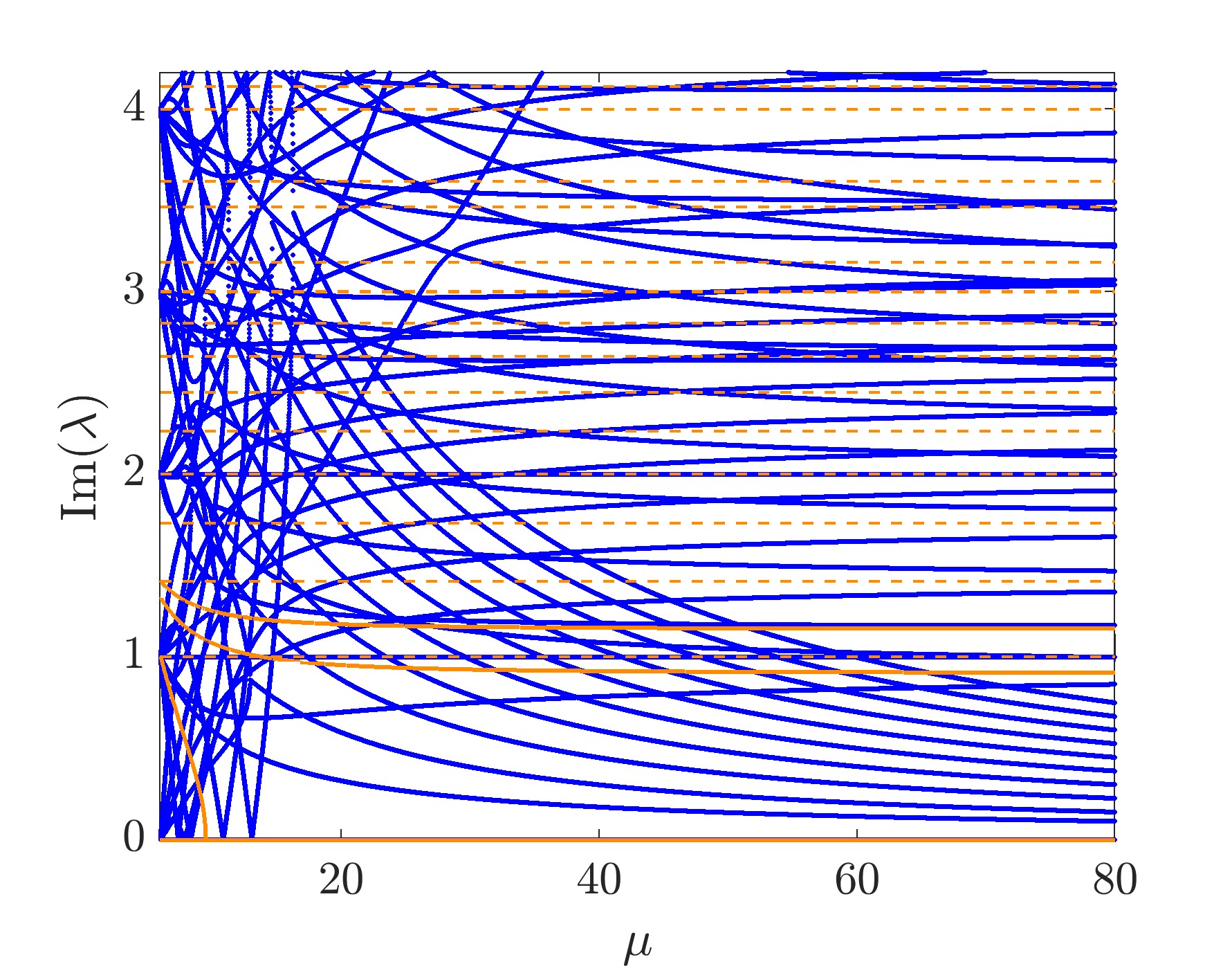}
\caption{(Color online) 
Top panels: the BdG spectra of the ground state GS, the single-charged
vortex state VX, the double-charged vortex state VX2, and the
triple-charged vortex VX3. Stable and unstable modes are depicted in
red and blue, respectively, and the horizontal dashed gold lines are the
asymptotic ground state modes. The former two states are spectrally
stable. Middle and bottom panels: the BdG spectra of the RDS, RDS+VX,
RDS+VX2, RDS+VX3 with central topological charges of $0, 1$, $2$, $3$,
respectively. The real and imaginary parts are plotted separately for
clarity. In addition to the ground state modes, there are RDS
modes. The horizontal solid (real) violet modes and (imaginary) gold
modes are, respectively, those of the solitary waves from the filament theory.
}
\label{RDS}
\end{figure*}

\subsubsection{Multiple rings}

The procedure for two rings readily generalizes in a
natural way to any number of rings. Skipping the details, the end-result is
the following formula for $N$ rings:
\[
\exp(-bl_{j})-\exp(-bl_{j+1})=ax_{j},\ \ \ j=1,2,\ldots, N.
\]
where $a,b$ are given in Eq.~(\ref{ab})
\begin{align}
r_{j}  &  =r_{0}+x_{j},\ \ \ j=1,2,\ldots, N;\nonumber\\
l_{j}  &  =x_{j}-x_{j-1},\ \ \ j=2,3,\ldots, N; \\
l_{1}  &  =l_{N+1}=\infty.\nonumber
\end{align}
For example, take $N=3$, $\mu=2$, and $\Omega=0.15$. Then, the above formulas 
yield $r_{0}=6.666$, $r_{1}=4.46$, $r_{2}=r_{0}$, and $r_{3}=8.87.$
Indeed, this again bears the imprint of the curvature and trap effects
balancing via $r_0$ and the RDS interactions and trap effects balancing via
the $x_j$'s. Here, too, the situation is reminiscent of the corresponding
multi-soliton generalization in the 1D case (where only the
inter-soliton interaction and the trap balance each other), as per the relevant
discussion, e.g., in the review~\cite{siambook}.


\section{Numerical Results}

Let us follow the same sequence as in the previous section for the analytical
considerations, but now for our numerical results. Namely, we start by
considering a single RDS state bearing a central vortex, and then move on to 
the multiple RDS configurations. 

\subsection{A single RDS with a central vortex}

First, some typical stationary states are shown in
Fig.~\ref{states}. For the single RDS states, the radii of the rings are indeed approximately independent of the charge $S$ in the TF regime, in line with the theoretical result. We defer to discuss their radii together with the multi-ring states for clarity and, for now, we focus on the spectra.

The BdG spectra of the ground state (GS) and the
single vortex (VX) state are shown in Fig.~\ref{RDS}. 
To our knowledge~\cite{fetter1,siambook}, these two
structures are likely the only fully robust states stemming from Eq.~(\ref{linearstates}) 
for all chemical potentials. The asymptotic ground state (TF background) excitations are well-known and were analyzed in, e.g., Ref.~\cite{siambook}. These modes (horizontal dashed gold lines) are in good agreement with the numerical results. In addition to these modes, the vortex spectrum bears an anomalous mode whose frequency tends to $0$ as $\mu \rightarrow \infty$. The latter mode corresponds
to the frequency of precession (around the center) that a vortex will
execute if displaced from the trap center~\cite{fetter1,siambook}.
It is well-known that vortex states of higher
vorticity are subject to instability~\cite{pubig} (see also~\cite{PW2}).
Indeed, the double-charge (VX2) and 
triple-charge (VX3) vortex spectra have a series of additional low-lying unstable bubbles, 
associated with complex eigenvalue quartets.
More generally, when the eigenvalue spectra are purely imaginary
(as for GS and VX), the waveform is spectrally stable, while
when it possesses eigenvalues with non-negative real
parts (as occurs in the bubbles for VX2 and VX3), it is spectrally
and dynamically unstable.

The results of the RDS series, containing a ring with an increasing
central vorticity, are summarized in the second and third rows of Fig.~\ref{RDS}.
The spectrum gets increasingly complex with increasing topological charge $S$. 
Here, we compare the full numerical results (in blue for the imaginary
parts and red for the real unstable ones) with those predicted by the
effective theory. It is important to recall that the latter has two
contributions. One comes from the corresponding state above without the ring, 
and the other concerns the point spectrum contributed by the vortex-RDS interaction, 
which is the one characterized by Eq.~(\ref{prediction}). While the horizontal dashed gold lines
in Fig.~\ref{RDS} represent the former, the horizontal solid gold and violet
lines of the effective particle picture developed above aspire to
capture the latter. 
The asymptotic behavior remains essentially the
same as the RDS, in line with the theoretical prediction that in the
large $\mu$ case, the topological charge effect disappears. Overall, it is found that the central 
vortex has a rather marginal effect towards stabilization. In fact, in
the higher-topological charge cases,
the vortex tends to introduce more instabilities due to the
contribution from the multiply charged vortex, i.e., the low-lying bubbles at the bottom of the 
states bearing topological charges $S \geq 2$, in line with the classic work of Ref.~\cite{pubig} (see 
also Ref.~\cite{carr}). 

Detailed inspection shows the following additional key features:
\begin{enumerate}
\item These states are stable with respect to perturbations along the modes $m=0$ and $1$ for the entire chemical potential domain studied.
\item The asymptotic instability growth rates (not counting the
  low-lying bubbles) in increasing order are due to the modes of
  $m=2$, $3$, $4$, and so on. 
  
\item The real RDS is exponentially unstable (purely real
eigenvalues), while all the vortical RDS+VX (enclosing a single-charge vortex)
and the higher-order RDS+VX2 and RDS+VX3 (enclosing respective higher charge 
vortices) are oscillatorily unstable (bearing complex eigenvalues).

\item We find two types of oscillatory instabilities
for the structures that include vortices:
\begin{itemize}
\item[(A)]
There is a ``cascade'' of instabilities from the RDS azimuthal Fourier modes
(here shown for relatively small values of $\mu$ as we are using
azimuthal modes $m=0,\dots,10$).
These instabilities emerge when two purely imaginary eigenvalues,
of opposite energies or Krein signature~\cite{siambook}, traveling in opposite directions 
(one up and one down) along the imaginary axis collide and create a
complex instability quartet. The growth rates (real parts) for these
instabilities monotonically increase and tend towards our particle picture predictions 
(see horizontal purple lines in Fig.~\ref{RDS}).
Indeed, this strongly suggests that the modes participating in these instabilities
are ones associated with azimuthal modulations $\propto e^{i n \theta}$, 
as per the filament theory analysis provided in Sec.~III.
This instability cascade is shared by all the configurations 
bearing vorticity in their center.
\item[(B)]
There is a series of instability ``bubbles''. In this case, a
negative Krein signature (see, in particular, the analysis
of Ref.~\cite{PW2}) purely complex eigenvalue 
(cf.~starting at $\lambda=4i$ for RDS+VX2)  travels up the imaginary
axis as $\mu$ increases
and collides successively with downward moving purely imaginary eigenvalues. These successive
collisions create small $\mu$-windows where the colliding eigenvalue pairs acquire a non-zero
real part and, after a relatively small increase in $\mu$, return back to the imaginary
axis (i.e., back to stabilization). This negative Krein signature
eigenvalue then continues its parkour up
along the imaginary axis and collides with another downward moving purely imaginary
eigenvalue creating yet another instability bubble and so on. 
This series of collisions seems to continue
indefinitely (at least for the ranges of $\mu$ that we have probed for $\mu\leq 80$).
We note that this type of instability bubbles is present for higher vortex charges
(i.e., for RDS+VX2 and RDS+VX3) but they are completely absent for the single-charge
case of RDS+VX, in line with what is known about the single VX configuration
vs.~VX2 or VX3, as is also shown in the top panels of Fig.~\ref{RDS}.
\end{itemize}

\end{enumerate}


Finally, we also briefly comment on the linear limit, recalling that this occurs
at small $\mu$, where the asymptotic theory of Eq.~(\ref{prediction}) is not applicable. 
As shown in earlier works, the dominant unstable mode near the linear limit may differ. In the 
neighborhood of the linear limit, the RDS is most unstable due to the $m=2$ mode, the RDS+VX due to 
the $m=3$ mode, and the RDS+VX2 due to the $m=4$ mode. However, we find this pattern does not persist 
for yet higher charges, e.g., the most unstable mode therein for the RDS+VX3 remains the $m=4$ mode. 
In addition, the presence of the charge in this regime may either accelerate or delay the 
instabilities depending on the modes. For example, in the RDS+VX spectrum, the $m=3$ mode 
destabilizes almost immediately, while the $m=2$ mode remains stable in this eigendirection 
up to a certain value of $\mu$.

\begin{figure}
 \includegraphics[height=8.4cm]{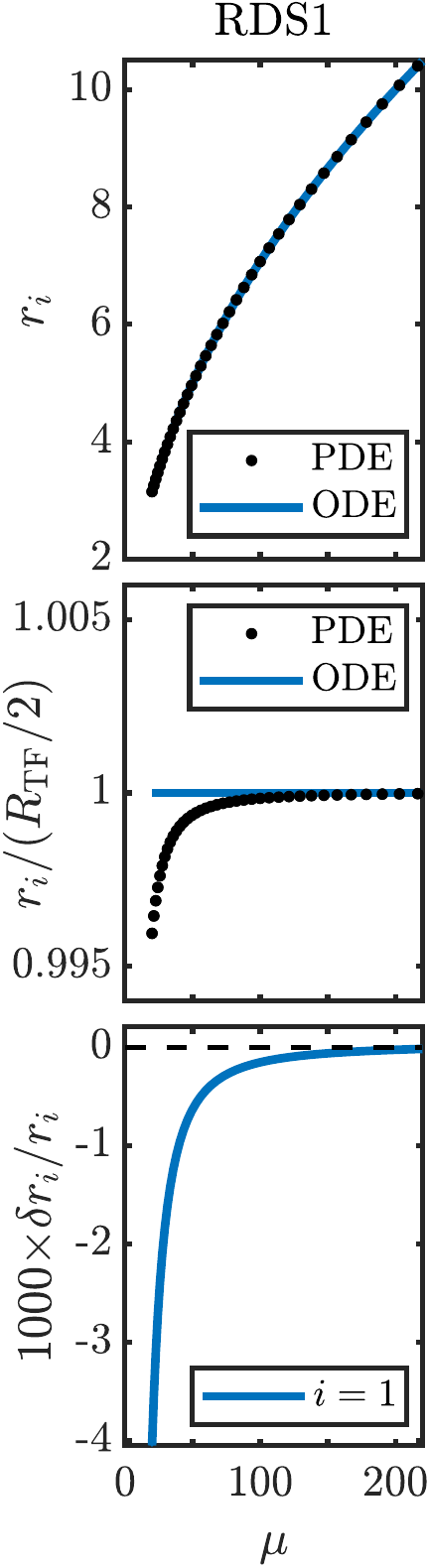} \!
 \includegraphics[height=8.4cm]{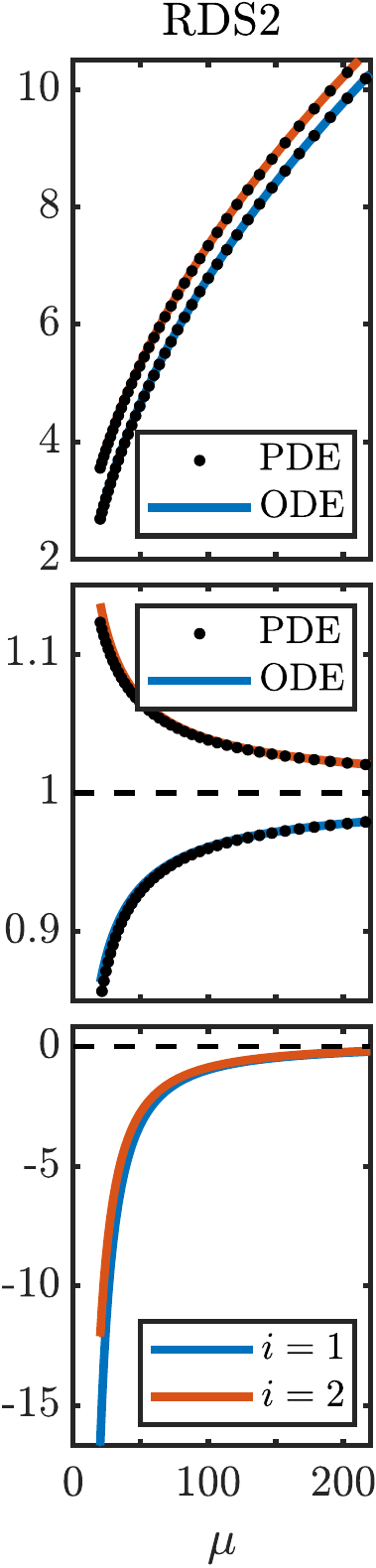} \!
 \includegraphics[height=8.4cm]{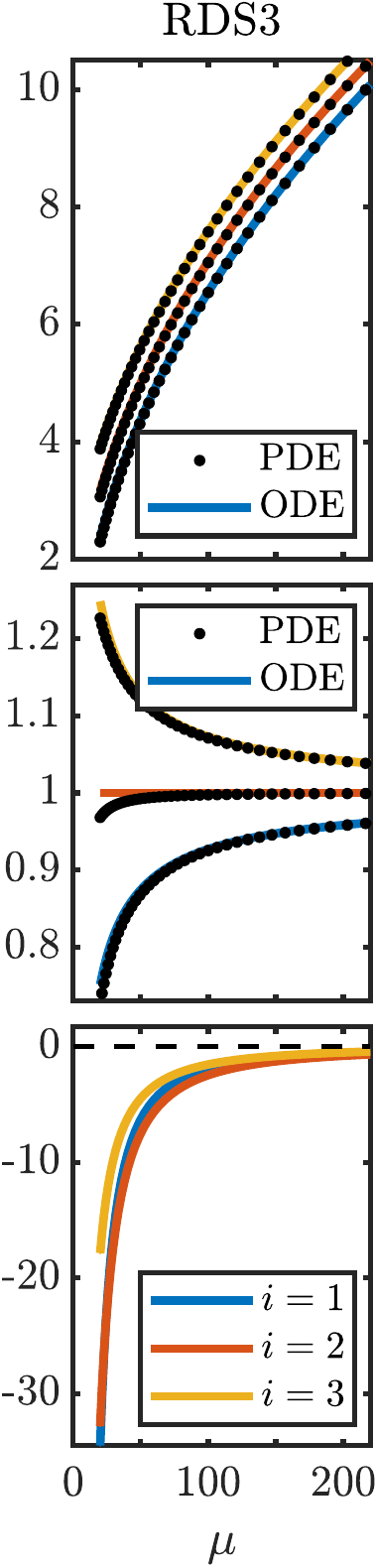} \!
 \includegraphics[height=8.4cm]{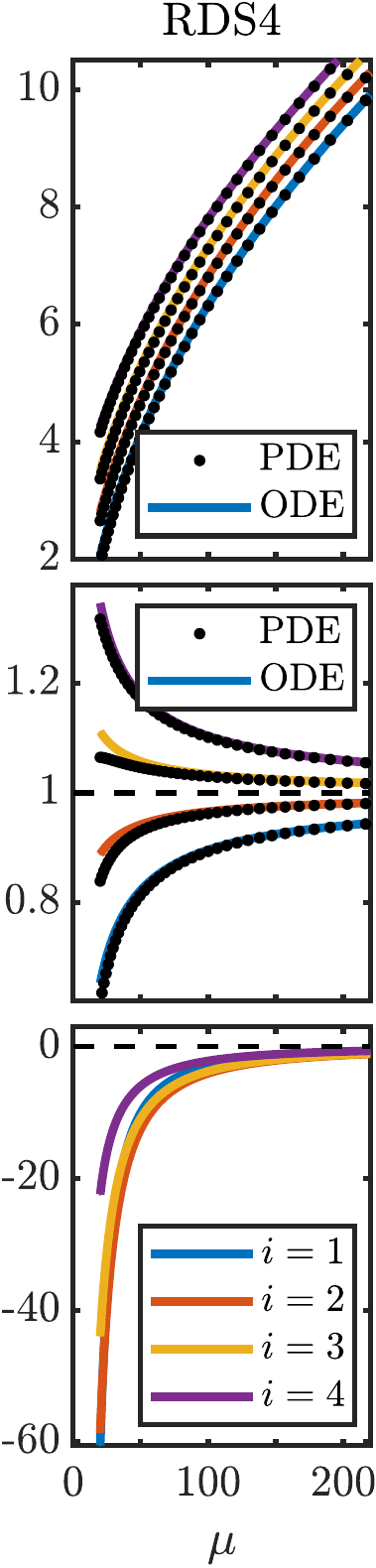} \!
\caption{(Color online)
Equilibrium RDS radii and the prediction using our particle theory. 
The different columns of panels depict, from left to right, the results
for the single (RDS1), double (RDS2), triple (RDS3) and quadruple (RDS4) RDS
in the absence of central vorticity.
The results including a central vortex are almost indistinguishable
from the ones depicted here and are thus not shown
(see Fig.~\ref{fig:plot_centrifugal} and its discussion in the text
for more details on the effect of the centrifugal potential
when central vortices are present).
%
The first row of panels depicts the RDS radii $r_i$ as a function $\mu$
from the numerically computed steady states of the original PDE (small black dots)
and from our theoretical prediction using the particle picture (solid lines)
for $\Omega=1$.
The second row of panels depicts the corresponding radii
rescaled by half of the TF radius $R_{\rm TF}/2=\sqrt{\mu/2}/\Omega$
[corresponding to the predicted location of the single RDS as per Eq.~(\ref{r0})].
The bottom row of panels depicts the corresponding relative errors between
the steady state radii and our prediction defined by 
$\delta r_i/r_i\equiv (r_{i,{\rm ODE}}- r_{i,{\rm PDE}})/r_{i,{\rm PDE}}$.
%
%
Note that for convenience, the relative errors have been multiplied by 1000,
yet they are never above 6\% for all the cases and chemical potential shown.
}
\label{RDSradius}
\end{figure}

\subsection{Multiple concentric RDS states}

\begin{figure}
\includegraphics[height=10.6cm]{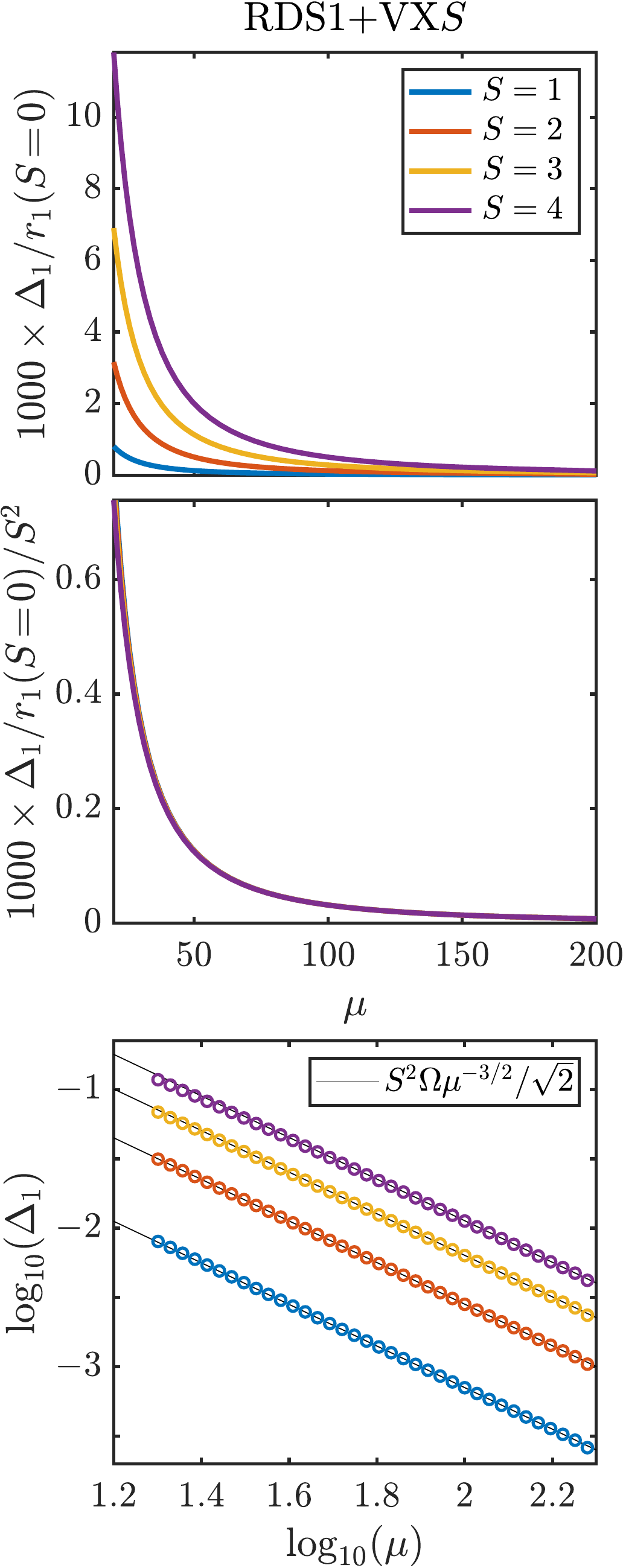}
\includegraphics[height=10.6cm]{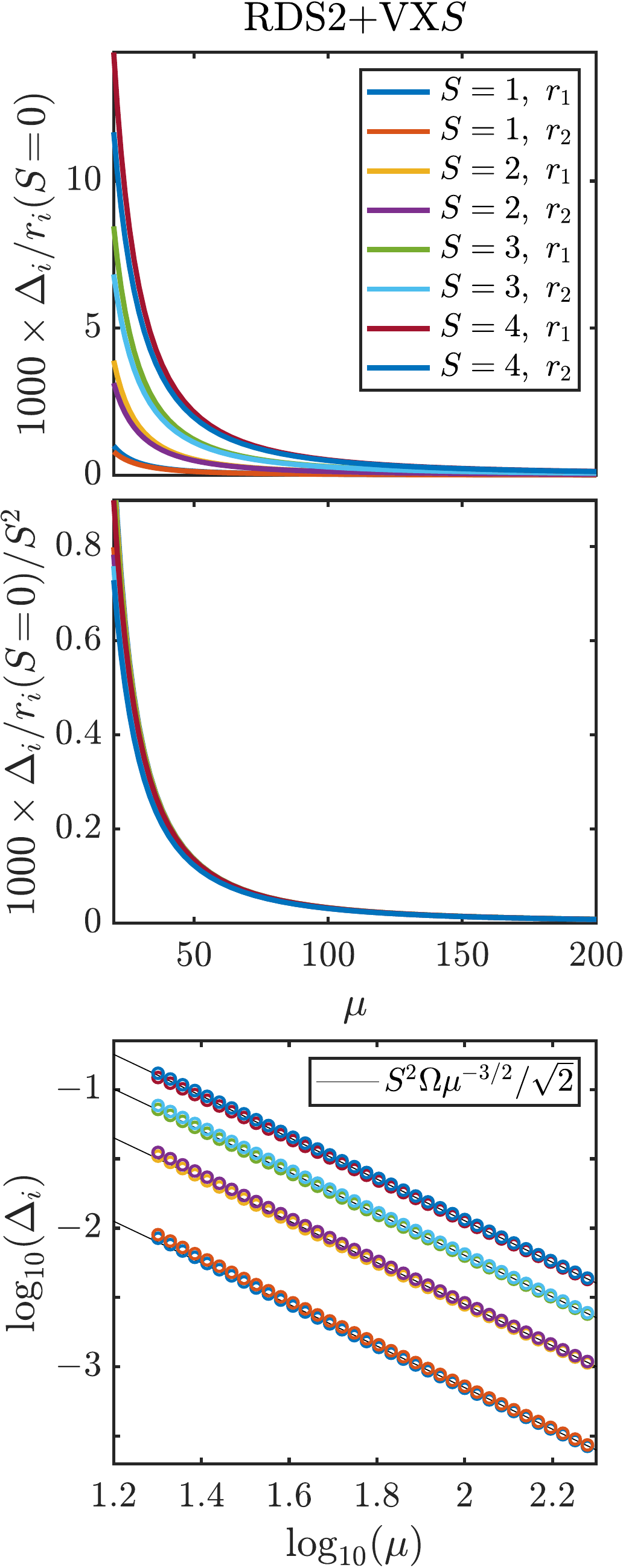}
\caption{(Color online)
Centrifugal effect on the ring positions as the central vortex charge $S$ is varied.
The left (right) column corresponds to the single (double) RDS.
The top panels depict the rings' displacement $\Delta_i=r_i(S)-r_i(S=0)$,
with respect to the case with no central vortex ($S=0$), when a central
vortex of charge $S=1,2,3,4$ is introduced.
The net effect of introducing a centrifugal potential through a central 
vortex is to sightly push out the rings' position.
By normalizing the displacement by $S^2$ all
the curves for the different values of $S$ overlap (see middle panels).
The bottom panels depict the rings displacement in log-lo scale showing 
the power-law decay of the centrifugal effect $\propto \mu^{-3/2}$;
see text for details.
Note that for convenience, in the first two rows of panels, 
the relative displacements have been multiplied by 1000.
}
\label{fig:plot_centrifugal}
\end{figure}

We now proceed with the ring radii of states involving multiple RDS.
The radii of the single and double rings 
at the linear limit are:
\begin{eqnarray}
r_{1\mathrm{linear}} &=& \sqrt{1+|n_{\theta}|}, \\
r_{2\mathrm{linear}} &=& \sqrt{|n_\theta|+2 \pm \sqrt{|n_\theta|+2}}.
\end{eqnarray}
The numerical results (not shown) are in good agreement with these predictions. In
the opposite TF regime, the particle picture radii compare favorably
with the numerical results, especially so as the TF limit is approached.
In the limit of very large $\mu$, all radii converge to 
$r_0=R_{\rm TF}/2$ [see Eq.~(\ref{r0})].
This is the case where the inter-soliton interaction
effect disappears. Nevertheless, the approach to the asymptotic
theory, already for $\mu \geq 30$, is clearly evident, as is the
ability of the theory to capture the relevant trend of the two-soliton radii
not only qualitatively but also quantitatively.
We have also extended the analyses to three and four rings, and again
the linear limit radii agree with theoretical results, and the analytical 
and numerical results closely match each other as we approach the TF regime.
In Fig.~\ref{RDSradius} we present the relevant results for the single, double,
triple, and quadruple RDS in the absence of central vorticity.
We note also that the corresponding results when a central vortex is included
are almost indistinguishable from the ones presented in Fig.~\ref{RDSradius}.
As we will see below (see Fig.~\ref{fig:plot_centrifugal} and its discussion),
the role of the effective centrifugal potential when central vortices are 
present is rather weak.
Figure~\ref{RDSradius} depicts the comparison of the RDS positions between the 
full numerical results and the results obtained from our particle picture theory.
The first two rows of panels in the figure depict the steady RDS radii as a 
function of $\mu$ for one to four rings.
It is interesting that the rings are approximately equally spaced,
featuring patterns reminiscent of the 1D equidistant lattices of dark
solitons, for the reasons revealed above, namely the pairwise balance of
interaction and trapping effects (around the equilibrium obtained from
the balance of curvature and trap effects).
%
The bottom row of panels in Fig.~\ref{RDSradius} depicts
the relative errors of the predicted ring positions.
As the figure suggests, the error on the prediction of the ring locations is 
small (note that these errors have been multiplied by 1000 for ease of presentation)
and further decreases as the chemical potential $\mu$ is increased. In fact,
for $\mu>30$, the ring positions prediction for all the studied cases
(namely, single, double, triple, and quadruple rings with and without vortices) 
have (relative) errors smaller than 1\%.
Therefore, as per the results depicted in Fig.~\ref{RDSradius}, not only are the
theoretical results in good qualitative agreement with the actual PDE results
as $\mu$ is varied, but more importantly, they are in excellent quantitative agreement
for moderate and large values of $\mu$.
%
%
%
For example, for the triple ring at $\mu=20$, the particle theory yields 
$(r_1,r_2=r_0,r_3)=(2.3756,3.1623,3.9490)$ while the full numerics yield
$(r_1,r_2,r_3)=(2.2964,3.0618,3.8798)$. The relative errors are (3.3\%, 3.1\%, 1.7\%),
in good agreement with the theoretical prediction. 
On the other hand, for the larger value of $\mu=100$, the particle theory yields 
$(r_1,r_2=r_0,r_3)=(6.5555,7.0711,7.5867)$ while the full numerics yield
$(r_1,r_2,r_3)=(6.5433,7.0538,7.5751)$. The relative errors are (0.18\%, 0.28\%, 0.15\%), 
in excellent agreement with the theoretical prediction.

\begin{figure*}
\includegraphics[width=0.24\textwidth]{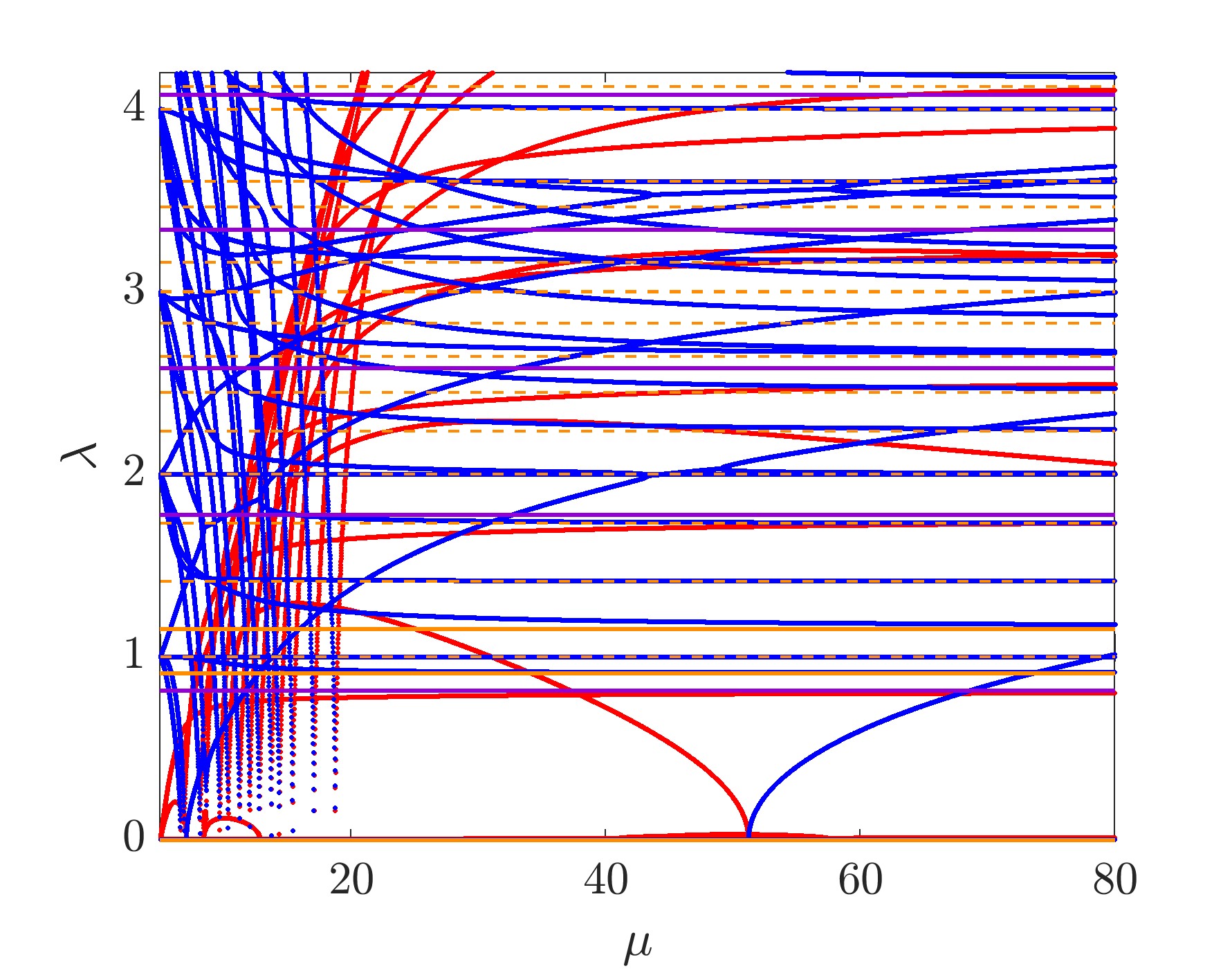}
\includegraphics[width=0.24\textwidth]{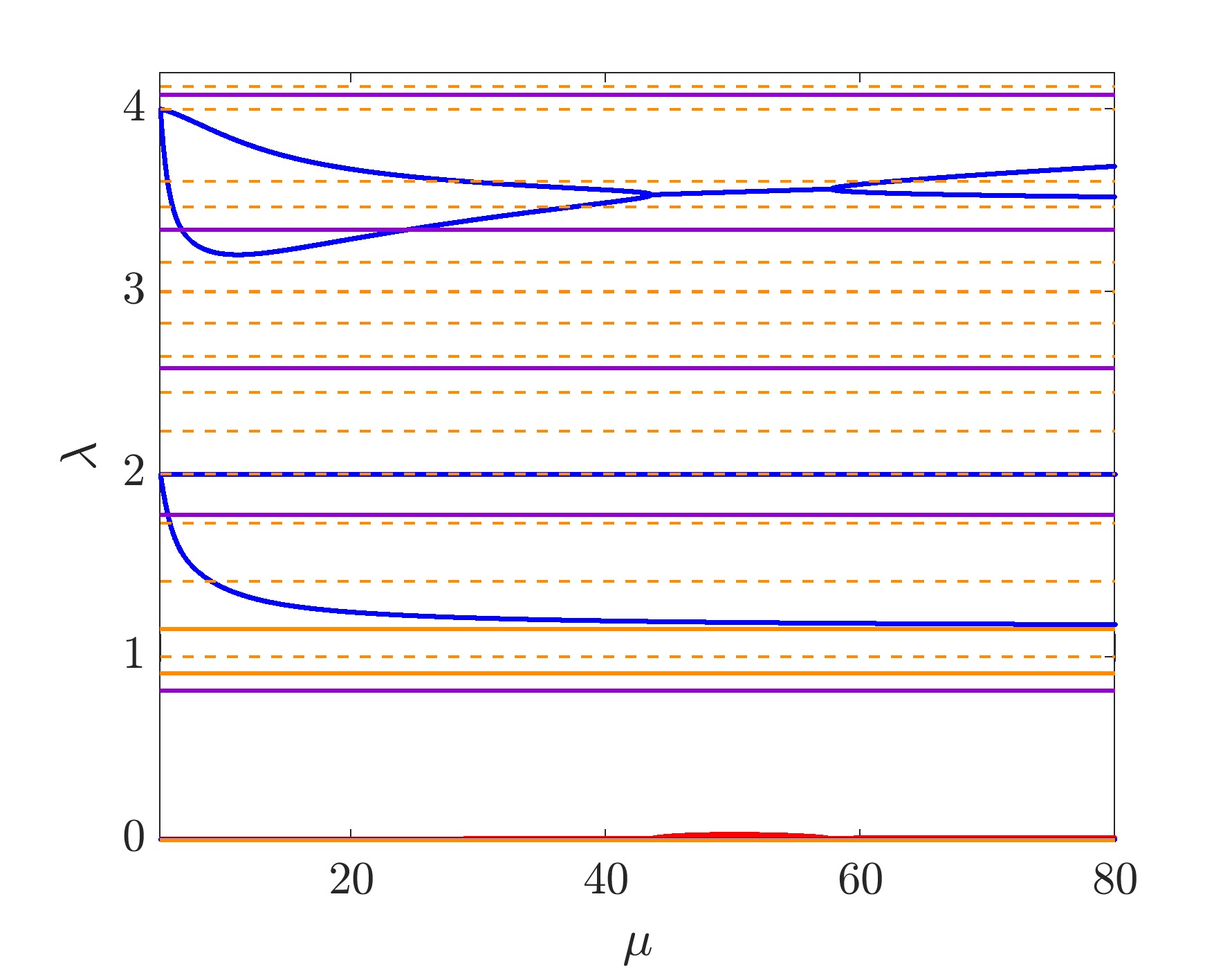}
\includegraphics[width=0.24\textwidth]{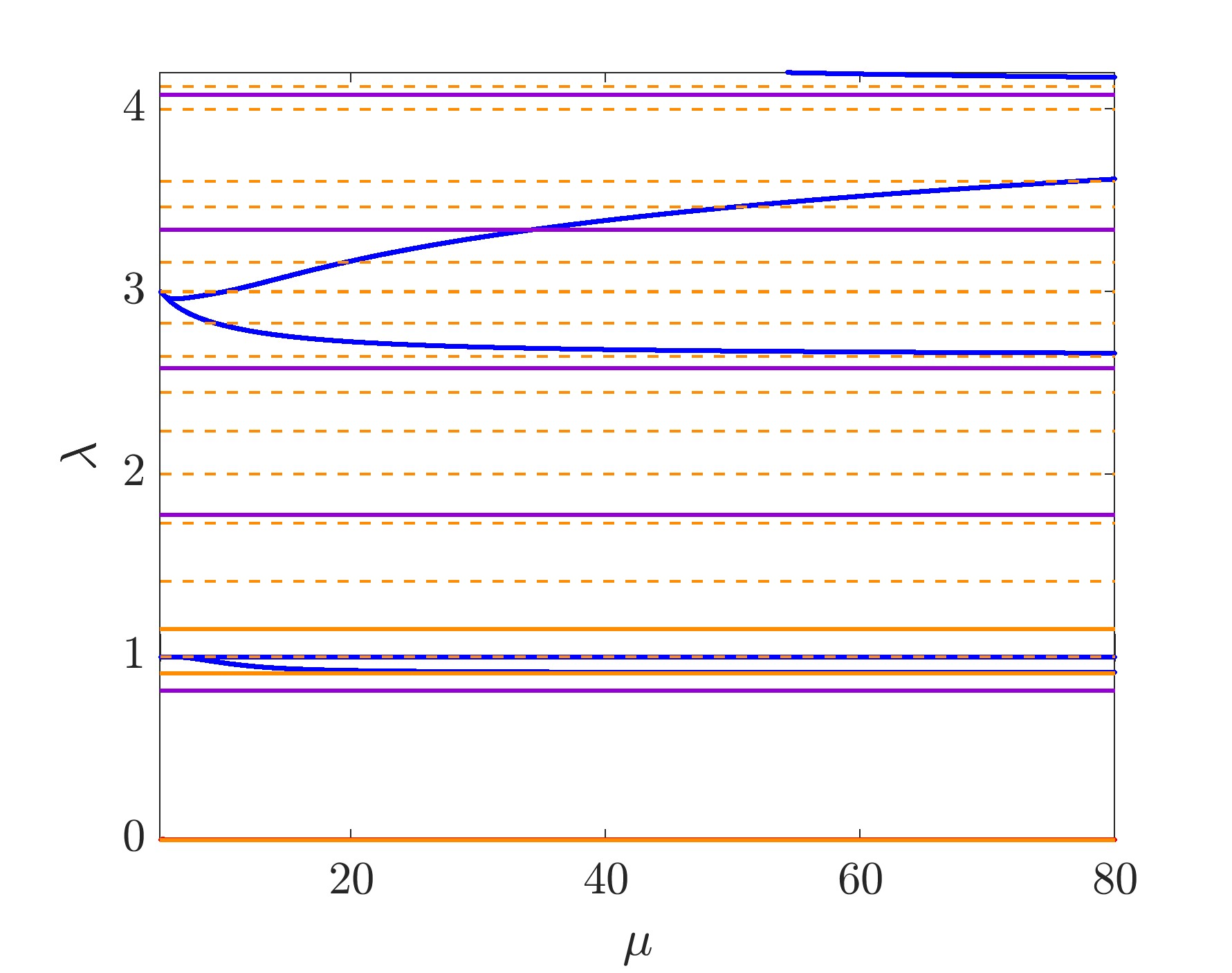}
\includegraphics[width=0.24\textwidth]{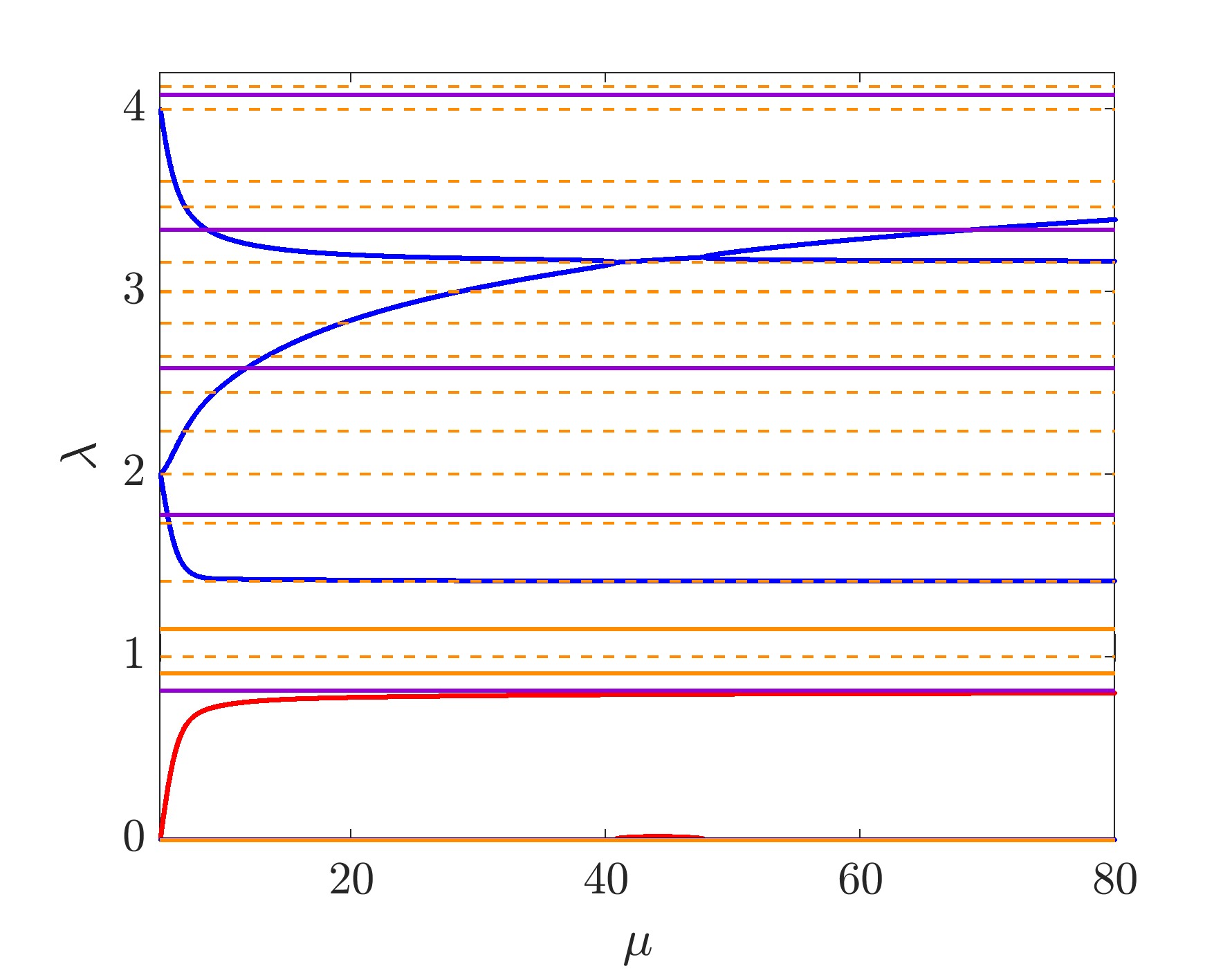}
\includegraphics[width=0.24\textwidth]{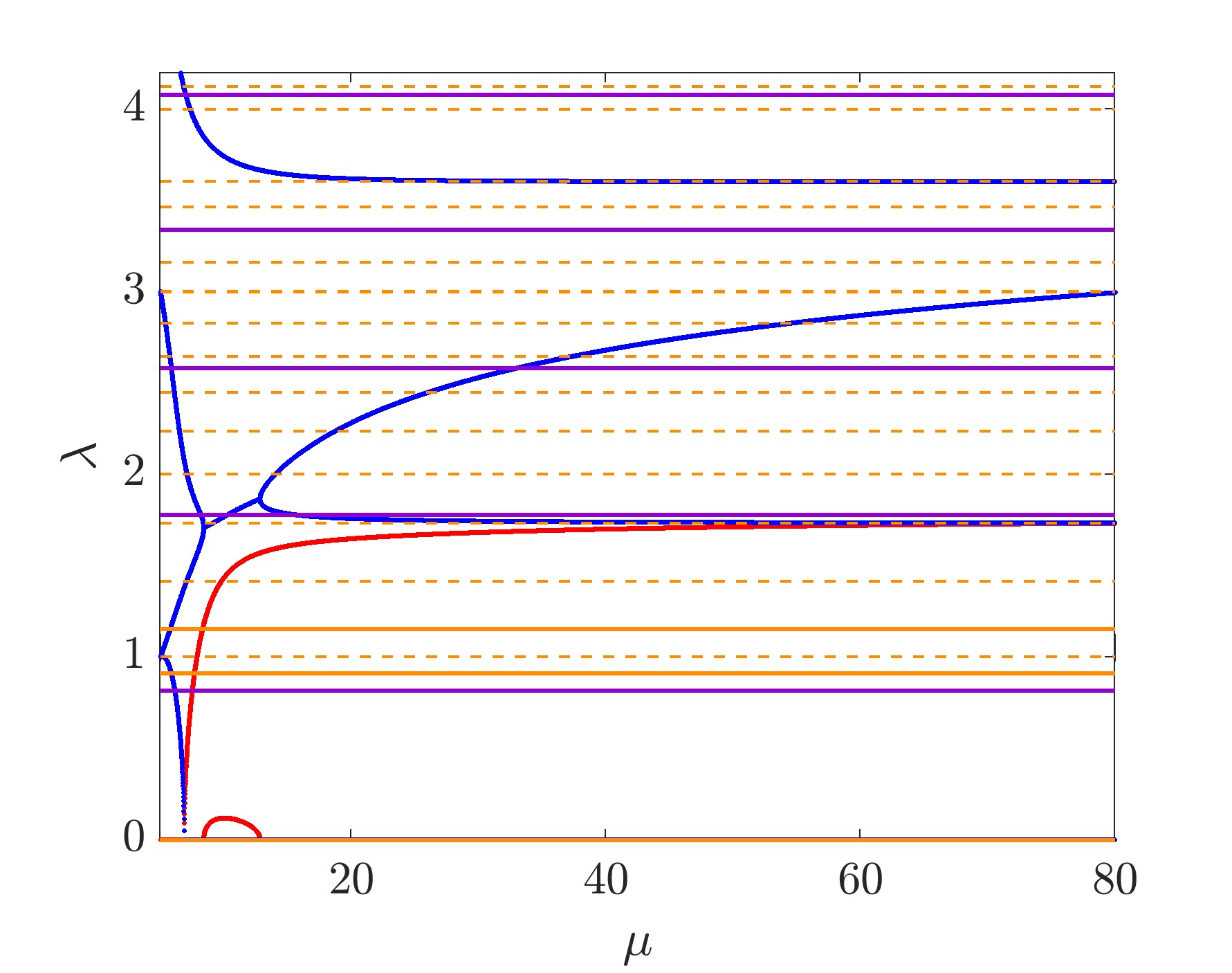}
\includegraphics[width=0.24\textwidth]{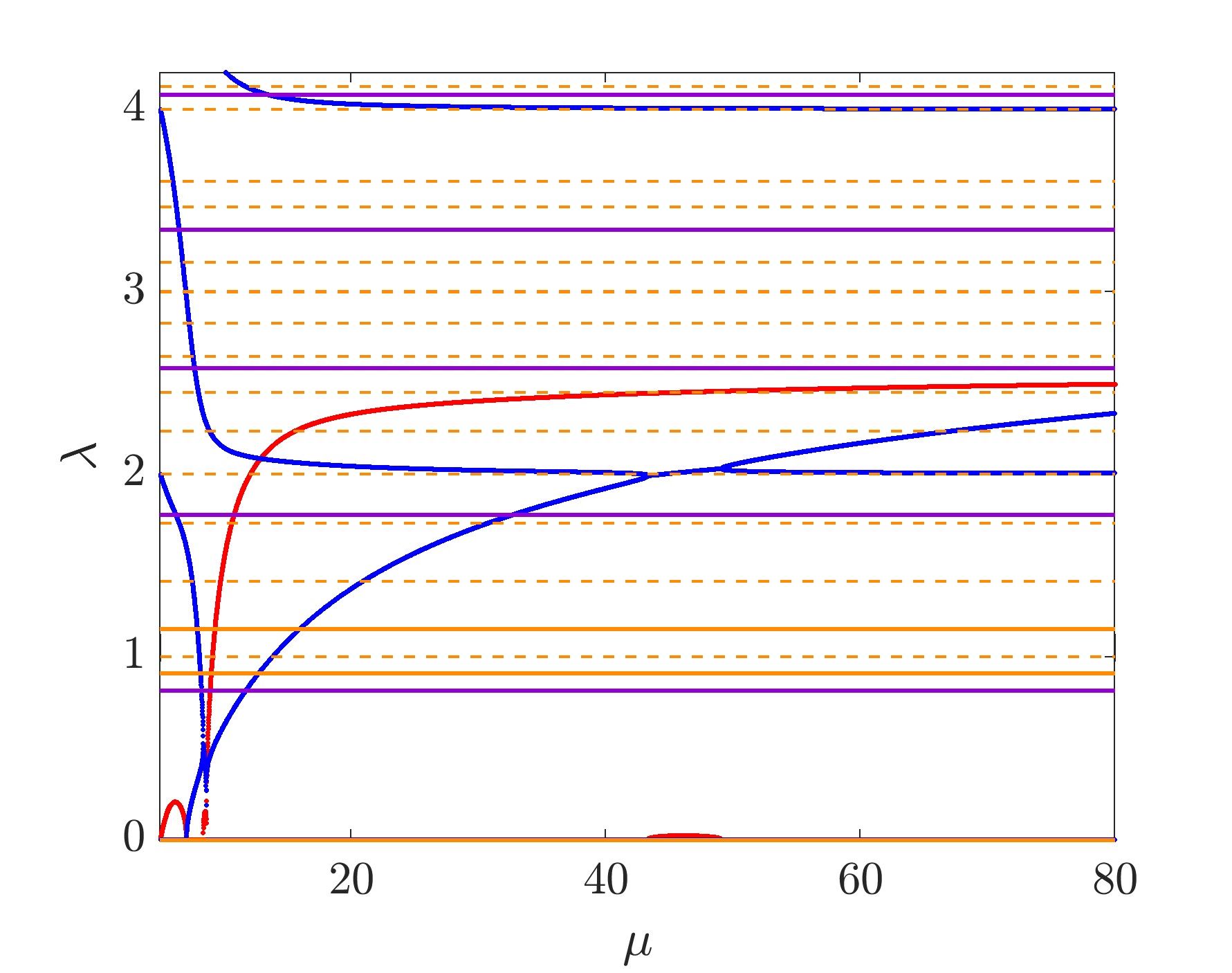}
\includegraphics[width=0.24\textwidth]{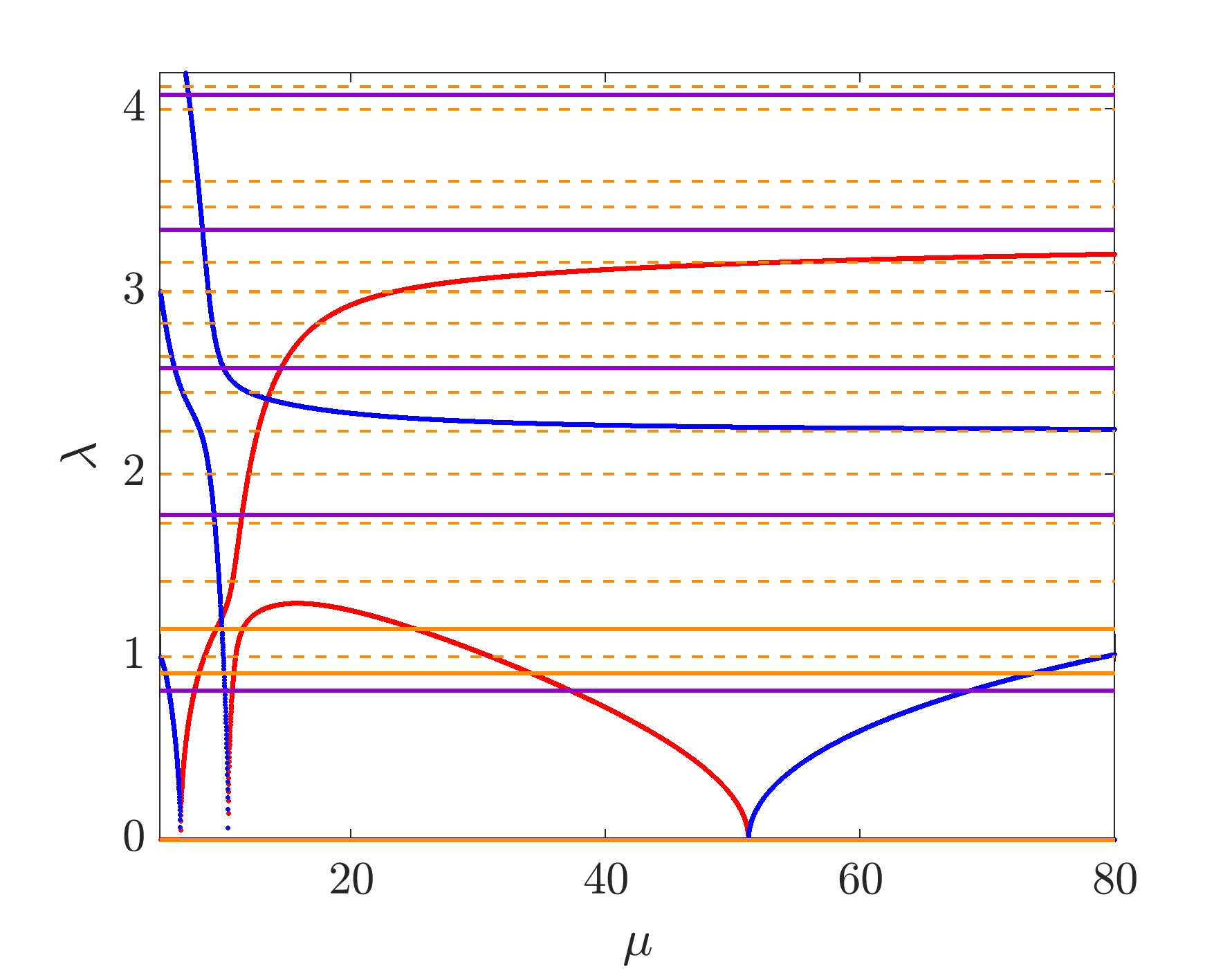}
\includegraphics[width=0.24\textwidth]{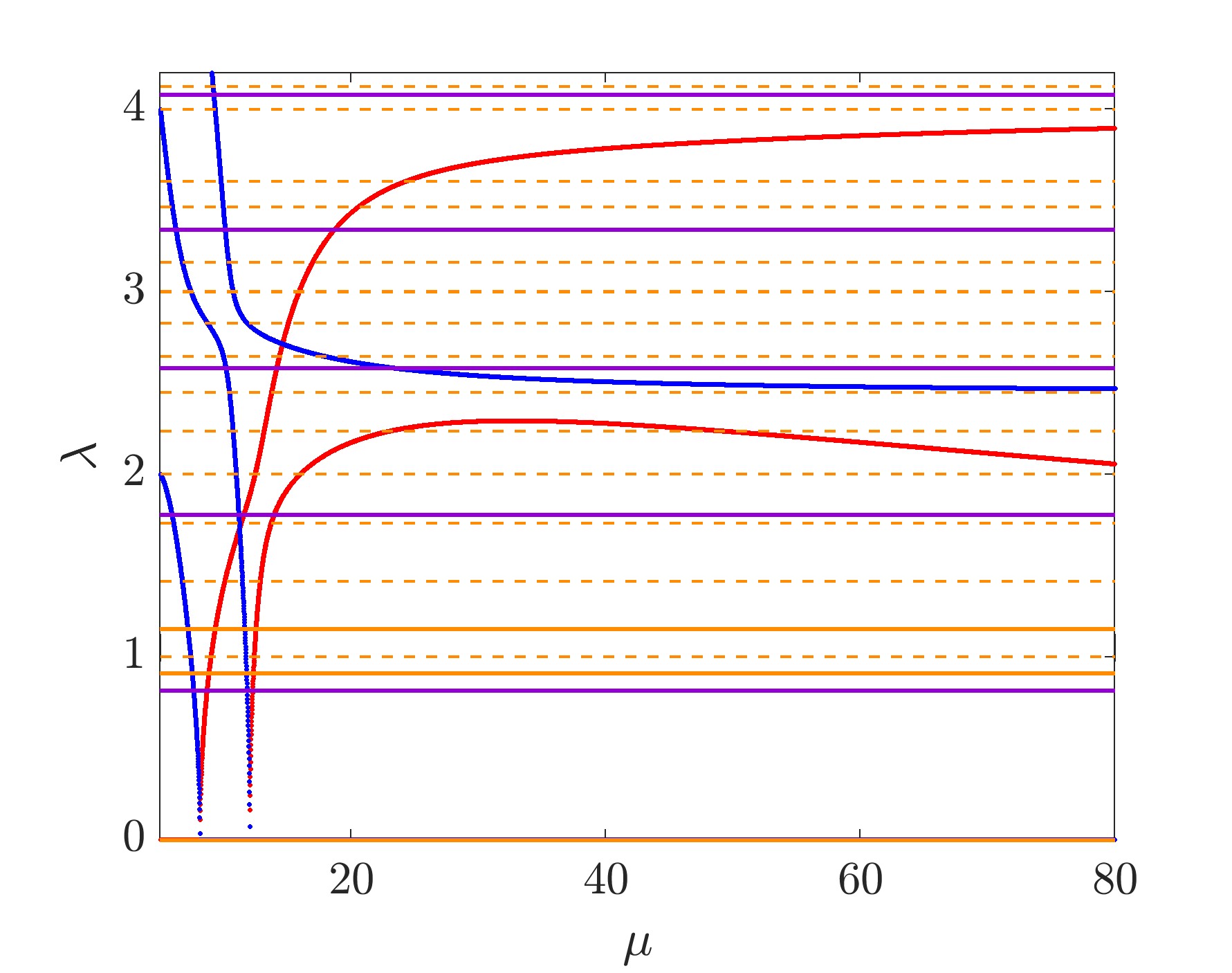}
\includegraphics[width=0.24\textwidth]{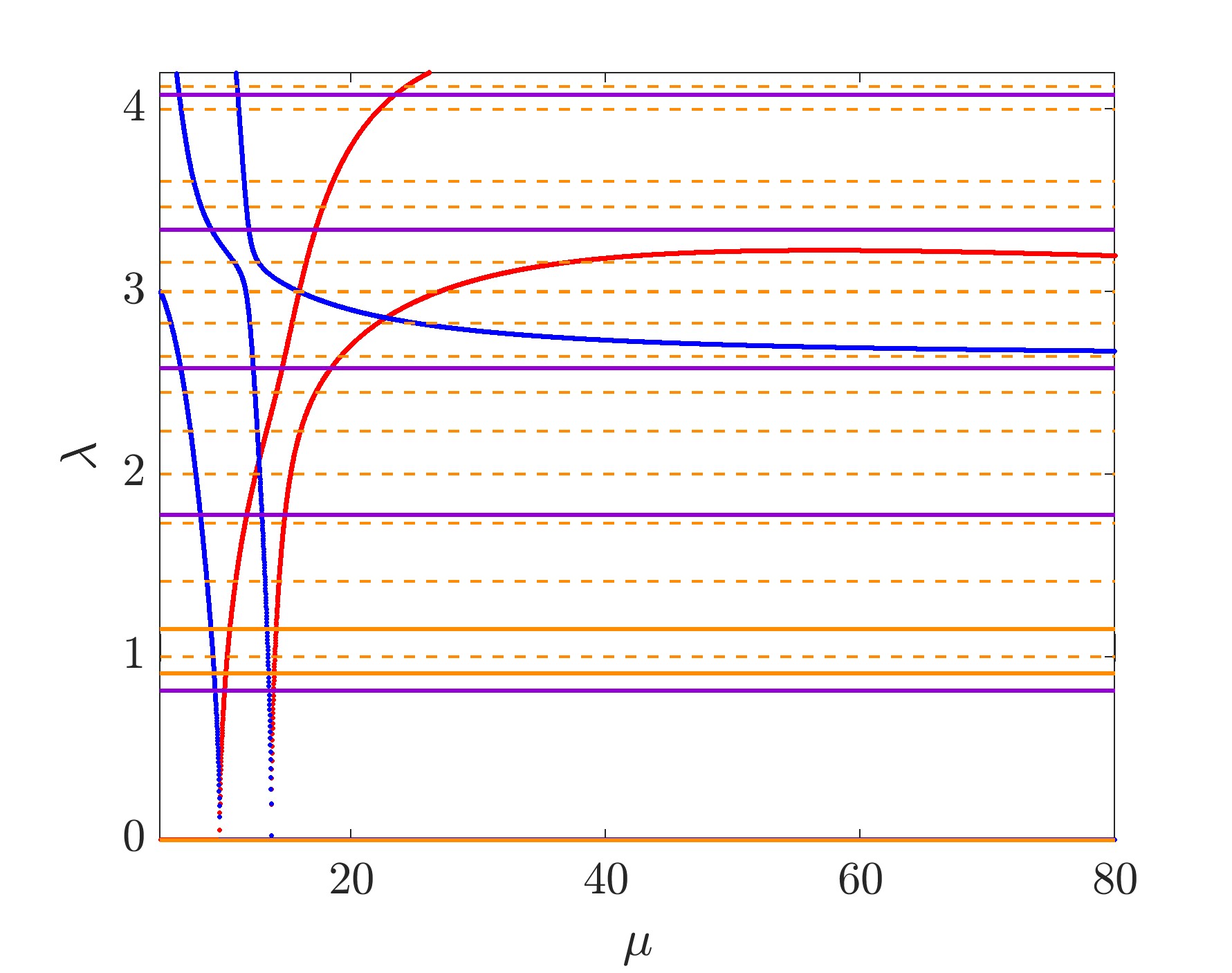}
\includegraphics[width=0.24\textwidth]{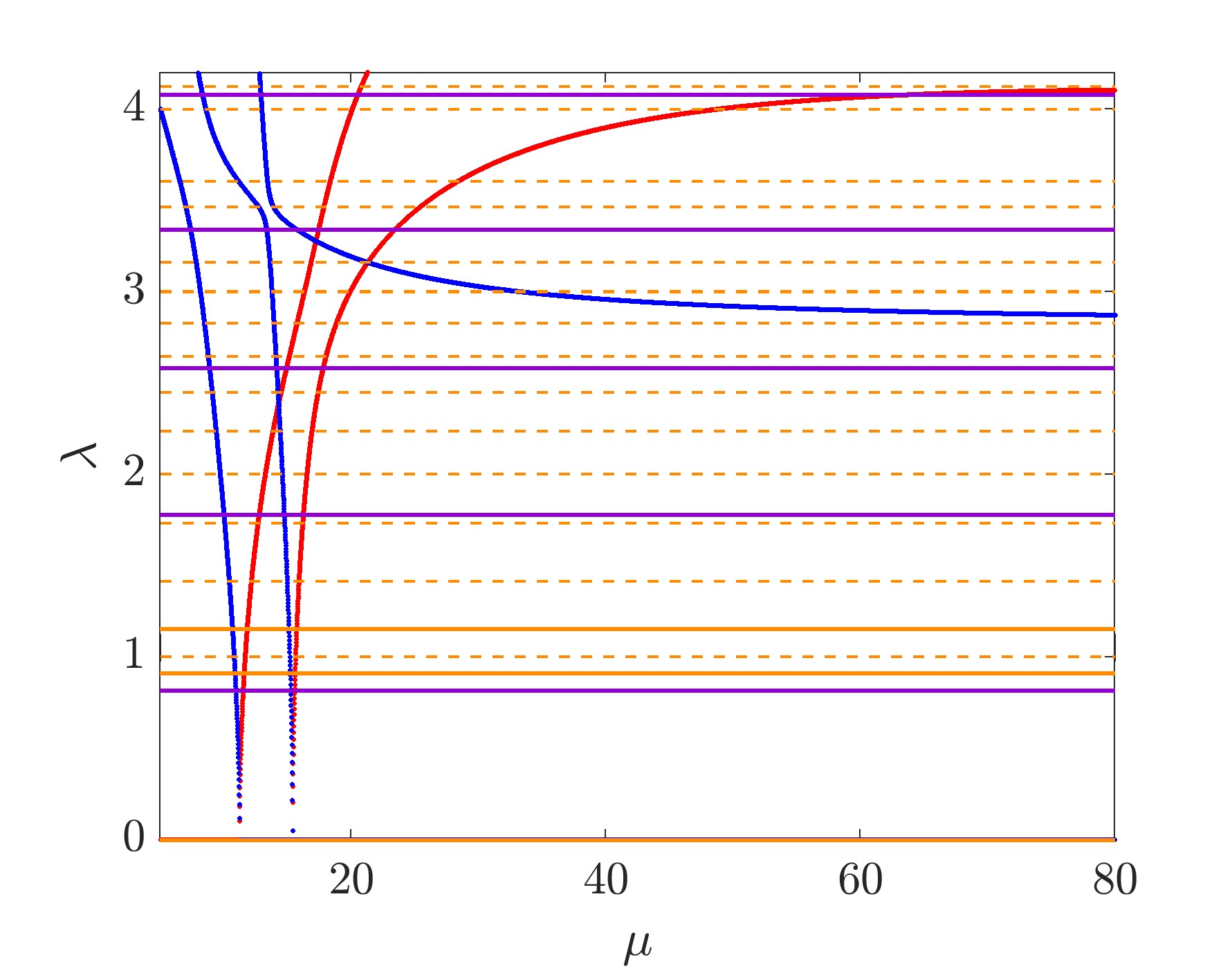}
\includegraphics[width=0.24\textwidth]{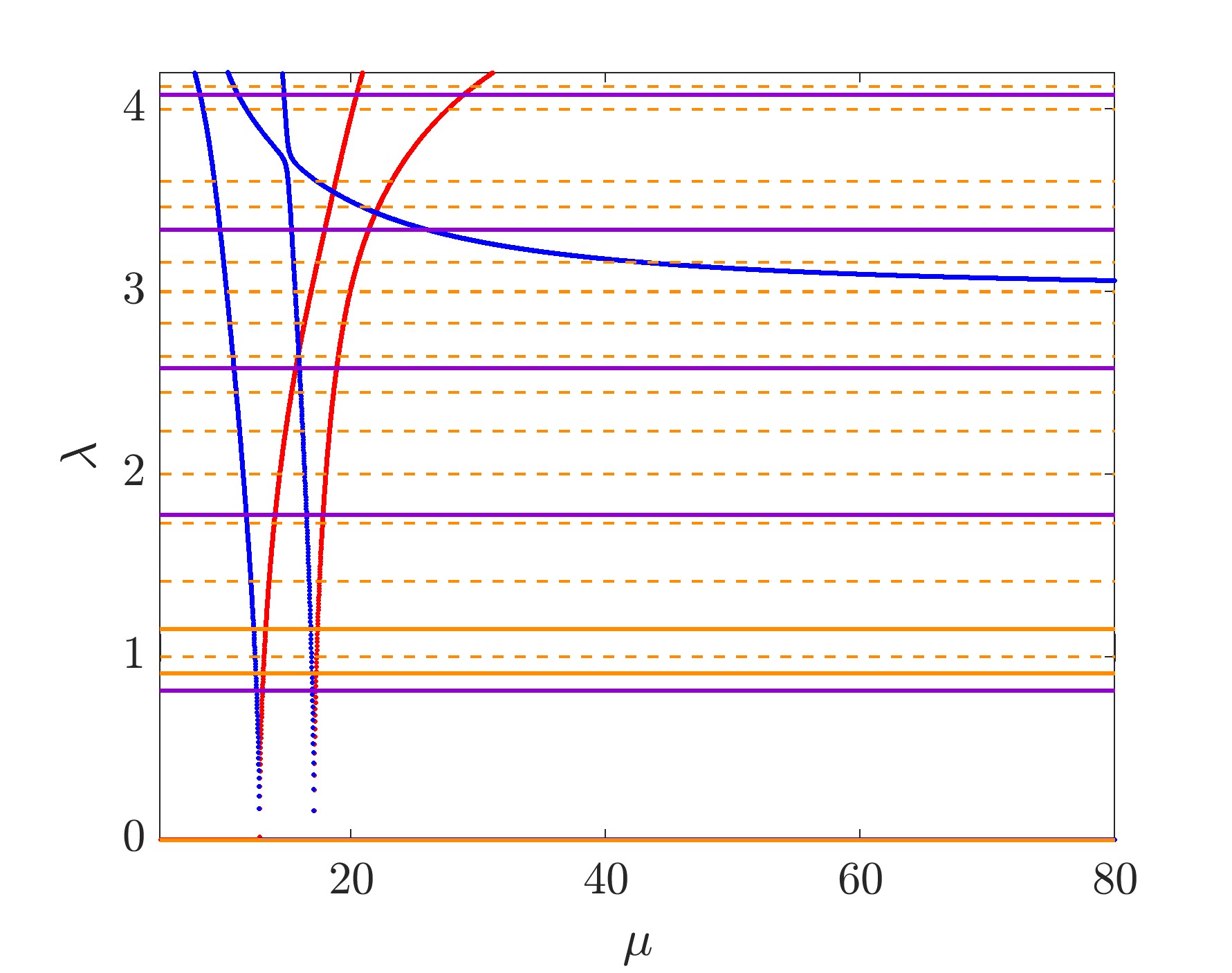}
\includegraphics[width=0.24\textwidth]{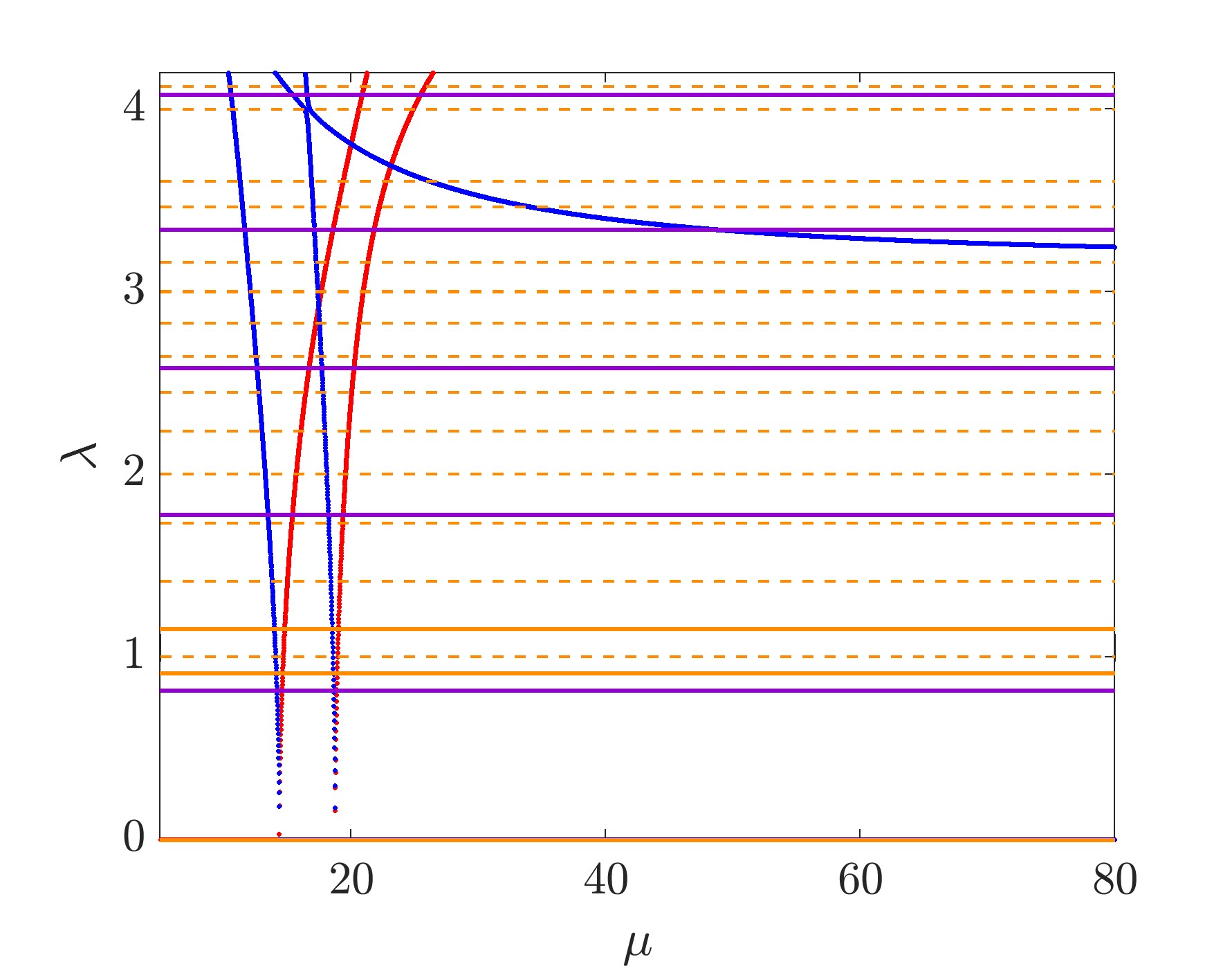}
\caption{(Color online)
The full double ring RDS2 spectrum (first panel) is quite complicated compared with the single RDS, therefore, the spectrum of modes $m=0, 1, ..., 10$ are in turn shown separately in the following panels. The RDS2 is not generically stable upon perturbation along the $m=0$ mode, but it is only unstable in a finite interval, i.e., the bubble, within our horizon. The single RDS asymptotic eigenvalues remain relevant, despite new modes emerge. 
}
\label{RDS2spectrum}
\end{figure*}

\begin{figure*}
\includegraphics[width=\textwidth]{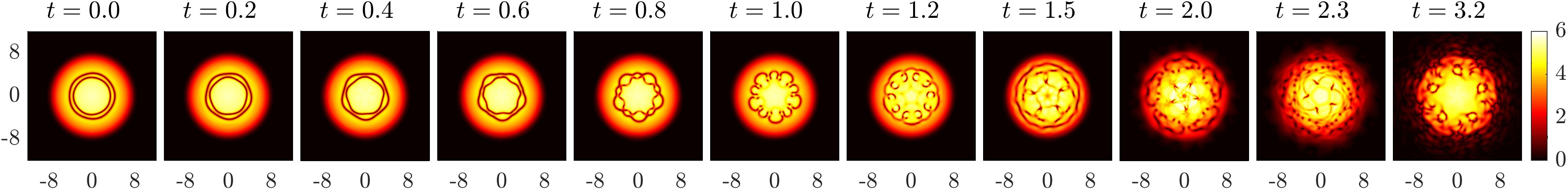}
\includegraphics[width=\textwidth]{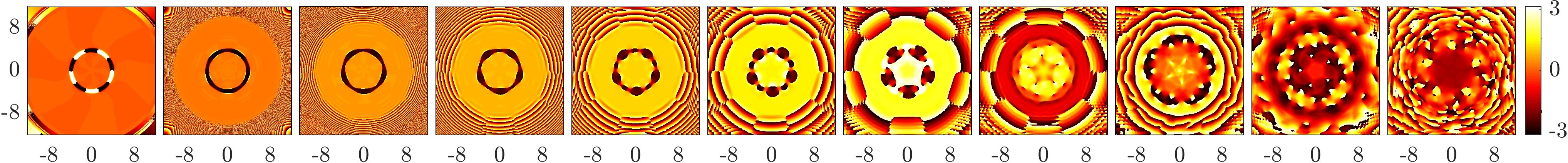}
\caption{(Color online) 
Destabilization dynamics for the double ring RDS2 along the $m=5$ eigenmode at $\mu=30$.
The initial state was taken to be a stationary double ring slightly (5\%) perturbed by the unstable
$m=5$ eigenmode.
For a movie of this destabilization dynamics please visit the following link:
\url{https://www.youtube.com/watch?v=mQs42F2bsHw}.
}
\label{fig:dyn_m5}
\end{figure*}

\begin{figure*}
\includegraphics[width=\textwidth]{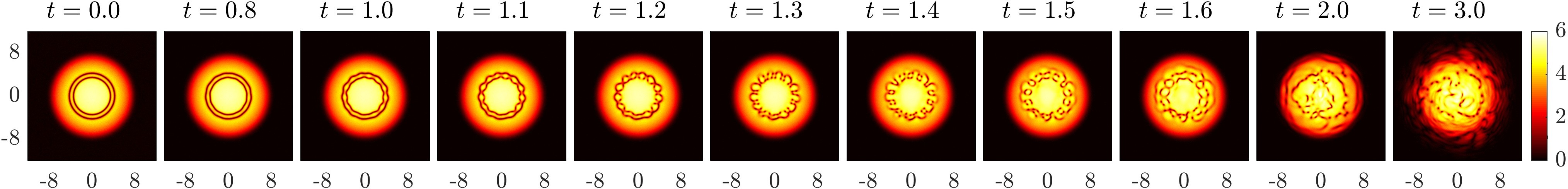}
\includegraphics[width=\textwidth]{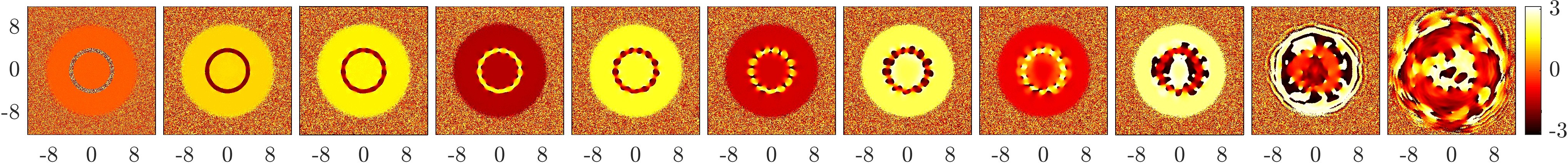}
\caption{(Color online) 
Destabilization dynamics for the double ring RDS2 for a random perturbation.
Similar to Fig.~\ref{fig:dyn_m5} but for a small random perturbation.
For a movie of this destabilization dynamics please visit the following link:
\url{https://www.youtube.com/watch?v=r0gcy0V2rwU}.
}
\label{fig:dyn_rand}
\end{figure*}

Finally, Fig.~\ref{fig:plot_centrifugal} depicts the effect of including
central vortices on the positions of the rings. The figure suggests that
the role of the effective centrifugal force induced by the central
vortex is to slightly increase the ring radii. 
In fact, for $\mu\simeq 20$ the relative displacement for a higher 
charge $S=4$ is only about 1\% and further decays for larger values of $\mu$.
As expected from the theory, the centrifugal effect is proportional to $S^2$
(note that $S$ appears as $S^2$ in the treatment of Sec.~III) as
evidenced in the middle panel of the figure where all displacement
curves coincide when rescaled by $S^2$.
Finally, the bottom panel in the figure shows that the effect of the centrifugal 
potential has a power-law decay as $\mu^{-3/2}$. In fact, it is straightforward
to check that, within our analytical considerations of Sec.~III, expanding
the expression for the rings displacement using the radii given in Eq.~(\ref{neweq1}) 
yields the (approximate) power-law decay for the displacement as 
$\Delta_i\simeq S^2 \Omega \mu^{-3/2}/\sqrt{2}$.
The numerical results for the displacement decay depicted in the bottom 
panels of the figure perfectly match this power-law decay (thin black lines).
%
%
Corresponding numerics as the ones depicted in Fig.~\ref{fig:plot_centrifugal} 
but for the triple and quadruple RDSs show equivalent results and thus, for brevity, 
are not shown here.

\begin{figure}
\includegraphics[width=0.9\columnwidth]{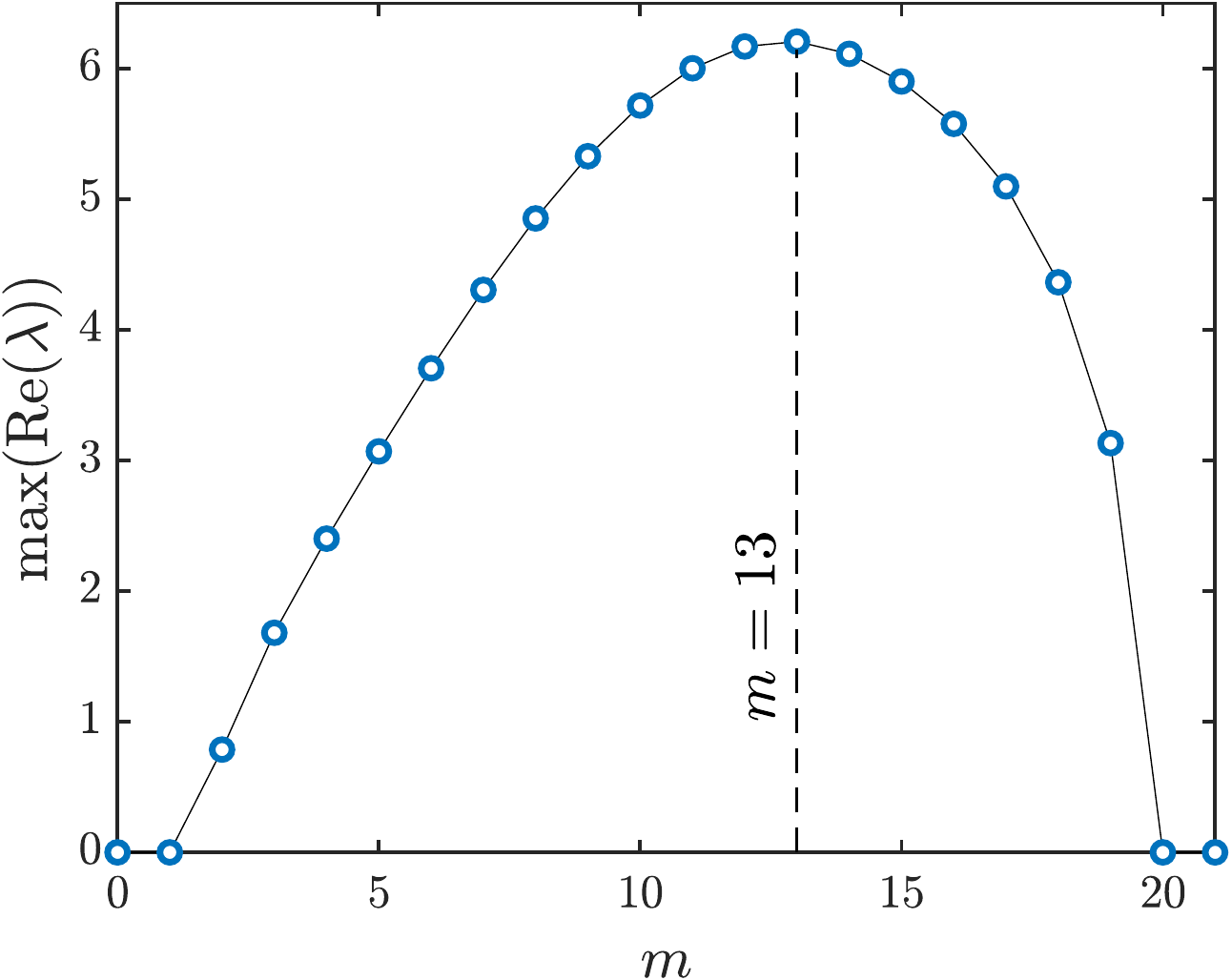}
\caption{
Mode instability for the double rings RDS2 for $\mu=30$.
Depicted is the maximal real part of the eigenvalue for azimuthal modes $m=0,\dots,21$.
}
\label{fig:RDS2_mu30_spectrum}
\end{figure}

Having studied the static properties of our particle prediction for the
single and multiple RDS configurations bearing (or not) central vortices, 
we now turn our attention to the stability of such structures.
%
Here, we focus on the RDS2 state for simplicity, and shall not explore all the 
structures systematically. Indeed, even this structure with two rings is 
significantly more complex than a single RDS. 
The corresponding spectrum is depicted in Fig.~\ref{RDS2spectrum}. 
The perturbation of the RDS2 along the mode $m=0$ is neither fully stable nor 
always unstable. Indeed, it appears to be oscillatory unstable in the interval 
$[43.6, 57.6]$ with a peak real maximum of only approximately $0.028$.
The RDS2 spectrum has an unusual feature around $\mu=50$ that two eigenvalues collide at $\lambda=0$ 
along the real axis and become purely imaginary. Detailed inspection shows that this instability is 
an $m=5$ mode, and the corresponding BdG eigenvector is radially concentrated around the two rings 
with the $\exp(i5\theta)$ phase signature. 
As Fig.~\ref{fig:dyn_m5} shows, upon perturbing the stationary state with this $m=5$ eigenvector 
(of approximately $5\%$ in norm, and the final state is renormalized to keep the norm unchanged), we 
find dynamically that the two rings deform towards two anti-aligned pentagons, i.e., the inner one 
is rotated with respect to the outer one by $\pi/5$. As the inner corners approach the outer edges, 
the rings nucleate numerous vortex pairs. 
The nucleation and the subsequent vortex dynamics depends on 
the specific chemical potential, and the number of vortices increases quite rapidly as the chemical 
potential increases within the regime of instability, but the decaying pathway and the subsequent 
$5$-fold symmetry are rather generic (in the direct numerical simulations that we conducted).
Finally, we show in Fig.~\ref{fig:dyn_rand} the destabilization of the double ring
starting from a small random perturbation. In this case the dynamics lacks the
$5$-fold symmetry of the previous figure as the destabilization occurs along a
competition of different modes (around $m=13$). 
In fact, in Fig.~\ref{fig:RDS2_mu30_spectrum}
we depict the instability growth rates for the different azimuthal $m$ modes. The figure
indicates a maximal instability growth rate for $m=13$ with its nearest neighbors
having similar, albeit slightly smaller, growth rates. This confirms the azimuthal modes
($m\sim 13$) selected in the destabilization of the randomly perturbed double ring 
depicted in Fig.~\ref{fig:dyn_rand}.

\subsection{Stabilization of the structures}

We will now investigate whether it is possible to stabilize the structures, 
so that they can be more readily created and studied in experiments. We use 
a radial Gaussian potential barrier, motivated by the earlier work
of Ref.~\cite{rcg:105}, but apply multiple ones, when needed, in a way
tailored to our problem. We
insert a barrier at each radial node, including the origin if a
central vortex is present there.
Nevertheless, it should be clarified that a radial node at the origin
is not always ``indispensable'' as a vortex pattern is not generically
unstable (in the TF limit). This touches upon the issue of persistent
currents discussed, e.g., in Ref.~\cite{rcg:99}.
More concretely, for a total of $k$ radial nodes, the additional perturbation potential reads:
\begin{eqnarray}
V_{\rm{pert}}=\sum_{i=1}^k A_i \exp[-(r-r_{0i})^2/(2\sigma_i^2)].
\end{eqnarray}
Here, $r_{0i}$ is the $i$th node (or barrier) location in the absence of
the perturbation, and $A_i$ and $\sigma_i$ are the corresponding strength
(amplitude) and width of the $i$th barrier, respectively. We work with the
simple but sufficient,
for our stabilization purposes, case where all the barriers have the
same amplitude $A$ and width $\sigma$. We have also studied a few cases
of unequal barriers, and find that the effects are qualitatively similar.

\begin{table}
\caption{
Prototypical parameter regimes where the pertinent solitary waves are
stable. The states are stable for the chosen $\mu$ and $\sigma$ in the
interval $A_{\mathrm{stable}}$, the stability at $A_{\mathrm{dy}}$ is
confirmed dynamically with direct numerical simulations
up to $t=1000$. Here, the growth rates typically get suppressed as $A$ grows until a state becomes fully stable, and it remains stable until $A$ is too large beyond which the soliton no longer exists.
\label{para}
}
\begin{tabular*}{0.98\columnwidth}{@{\extracolsep{\fill}} l c c c c}
\hline
\hline
~States &$\mu$ &$\sigma$ &$A_{\mathrm{stable}}$ &$A_{\mathrm{dy}}$ \\
\hline
~RDS+VX  &$5.5$ &$0.5$ & $[1.7, 3.8]$ &$2.0$ \\
\hline
~RDS+VX2  &$6.0$ &$0.5$ & $[0.6, 2.8]$ &$1.0$ \\
\hline
~RDS+VX3  &$7.0$ &$0.5$ & $[1.5, 3.0]$ &$2.0$ \\
\hline
~RDS2  &$6.3$ &$0.4$ & $[1.9, 3.0]$ &$2.0$ \\
\hline
~RDS3  &$8.2$ &$0.3$ & $[2.1, 3.1]$ &$2.4$ \\
\hline
~RDS4  &$10.2$ &$0.25$ & $[2.2, 3.3]$ &$2.4$ \\
\hline
\hline
\end{tabular*}
\end{table}

We first choose a typical state close to the linear limit such that it
is neither too close to the linear limit, nor it already possesses
numerous unstable modes (as in the TF limit); in other words, we focus on 
a regime of intermediate chemical potentials. 
Then, we fix a reasonable width $\sigma$ and scan the strength $A$,
starting from $0$ to a threshold $A_f$ beyond which the soliton ceases
to exist. We find that $\sigma$ appears to work best if it is
reasonably wide but nearest barriers do not significantly overlap,
i.e., sharp and strong barriers interestingly do not appear to be particularly
efficient towards the stabilization of these structures. Nevertheless,
with these guiding rules, it is typically not difficult to
stabilize the structures, i.e., there are presumably wide ranges of
parameters in the $(A,\sigma)$ parameter space which feature spectral stability.
Indeed, we have stabilized all of the structures considered with our
simple perturbation potential. Considering the large number of solitary waves we are
studying here, we present some typical parameters where these solitary
waves are fully stable to illustrate the proof-of-principle,
and we shall not explore here the full $(\mu, A, \sigma)$ parameter
space further. The results are summarized in Table~\ref{para}. The
stability properties of these waves are confirmed dynamically for each 
state up to as large as $t=1000$ (results not shown here).

\section{Conclusions and Future Challenges}

In the present work, we have revisited a widely studied structure in
the form of the ring dark soliton (RDS). Motivated, in part, by
earlier explorations in nonlinear optics~\cite{nepom1,nepom2}
(and associated optical experiments) of vortex-stripe interactions
and the corresponding nonlinear Aharonov-Bohm scattering
considerations, but also by much more recent considerations of multi-ring
structures constructed in single- as well as in multiple-component 
atomic BECs~\cite{bigelow}, we have explored both classes of such configurations.
Namely, we have examined RDS structures harboring vortices
at their centers, as well as multiple RDS configurations interacting
with each other, while sustaining effects of curvature and of the
external parabolic trap, as is typically relevant to the BEC setting.

We have found that, generically, the composite configurations involving a RDS
are unstable. While vortices appear to have a stabilizing effect,
this effect is 
too weak in the Thomas-Fermi large density limit in which we could analytically 
quantify it. The multi-ring configurations were found to naturally generalize
the 1D ordered (equidistant) lattices of dark solitons.
Here, we could reveal the balance of the different ``forces'' on the
rings, namely the balance between curvature pushing the rings
outwards and the trap restoring them towards its center, as well
as the quasi-1D balance between the inter-soliton interaction and the
above two features. The interplay 
of these different effects leads to the existence of steady states which we 
could quantify in good agreement with numerical computations. Nevertheless,
these structures were also typically found to be unstable with the
corresponding instability dynamics leading to the (transient) formation of
complex multi-vortex-lattice configurations.
Interestingly, in all the cases of the different nonlinear waves,
stabilization could be achieved by the presence of external radial
potentials. This feature was found to be generic for all the states considered
herein.

These findings suggest a number of possibilities for future
efforts. It is well-known that 3D generalizations of RDS
exist but are also dynamically unstable~\cite{babev}. Hence,
it is perhaps of more interest to examine the interactions
of vortex rings with vortex lines~\cite{bisset} and to analyze the relevant
particle/filament picture for which we are not aware of any
theoretical results.
In the same spirit as our multi-ring structures herein,
one can also explore multi-vortex-ring entities and lattices thereof
in 3D BECs, which are also of interest in their own right~\cite{danaila04}.
Such studies will be reported in future publications.

\section*{Acknowledgments}
W.W. acknowledges support from the National Science Foundation of China under Grant No. 12004268, the Fundamental Research Funds for the Central Universities, China, and the Science Speciality Program of Sichuan University under Grant No.~2020SCUNL210.
R.C.G.~acknowledges support from the US National Science Foundation
under Grant No.~PHY-1603058.
P.G.K.~acknowledges support from the US National Science Foundation
under Grants No.~PHY-1602994 and DMS-1809074. He also acknowledges support from the Leverhulme Trust via a Visiting Fellowship and the Mathematical Institute of the University of Oxford for its hospitality during part of this work. We thank the Emei cluster at Sichuan university for providing HPC resources.


\begin{thebibliography}{99}

\bibitem{review_dalfovo}
F. Dalfovo, S. Giorgini, L.P. Pitaevskii and S. Stringari,
Theory of Bose-Einstein condensation in trapped gases,
Rev.\ Mod.\ Phys.\ {\bf 71} (1999) 463--512.

\bibitem{becbook1}
C.J. Pethick and H. Smith,
{\it Bose-Einstein condensation in dilute gases},
(Cambridge University Press, Cambridge, 2002).

\bibitem{becbook2}
L.~P.~Pitaevskii and S.~Stringari,
{\it Bose-Einstein Condensation and Superfluidity},
Oxford University Press (Oxford, 2016).


\bibitem{kivshar_agr} 
Yu.S. Kivshar and G.P. Agrawal,
{\sl Optical solitons: from fibers to photonic crystals},
Academic Press (San Diego, 2003).

\bibitem{Infeld} E. Infeld and G. Rowlands, {\sl Nonlinear Waves, Solitons
and Chaos} (Cambridge University Press, 1990).

\bibitem{abl3} M.J. Ablowitz,
{\it Nonlinear Dispersive Waves},
Cambridge University Press (Cambridge, 2011).

\bibitem{chap01:becgss}
A. Griffin, D.W. Snoke, and S. Stringari (eds.),
{\it Bose-Einstein Condensation}
(Cambridge University Press, Cambridge, 1995).

\bibitem{chap01:anderson}
M.H.J. Anderson, J.R. Ensher, M.R. Matthews, C.E. Wieman,
Observation of Bose-Einstein condensation in a dilute atomic vapor,
Science {\bf 269} (1995) 198--201.

\bibitem{chap01:davis}
K.B. Davis, M.-O. Mewes, M.R. Andrews, N.J. van Druten,
D.S. Durfee, D.M. Kurn and W. Ketterle,
Bose-Einstein condensation in a gas of sodium atoms,
Phys.\ Rev.\ Lett.\ {\bf 75} (1995) 3969--3973.

\bibitem{chap01:bradley}
C.C. Bradley, C.A. Sackett, J.J. Tollett, and R.G. Hulet,
Evidence of Bose-Einstein condensation in an atomic gas with
attractive interactions,
Phys.\ Rev.\ Lett.\ {\bf 75} (1995) 1687--1690.

\bibitem{chap01:rmpnlA}
E. A. Cornell and C. E. Wieman,
Bose-Einstein condensation in a dilute gas, the first 70 years and some recent experiments,
Rev.\ Mod.\ Phys.\ {\bf 74} (2002) 875--893.

\bibitem{chap01:rmpnlB}
W. Ketterle,
When atoms behave as waves: Bose-Einstein condensation and the atom laser,
Rev.\ Mod.\ Phys.\ {\bf 74} (2002) 1131--1151.

\bibitem{Matthews99}
M. R.~Matthews, B. P.~Anderson, P. C.~Haljan, D. S.~Hall,
C. E.~Wieman, and E. A.~Cornell,
Vortices in a Bose-Einstein Condensate,
Phys.\ Rev.\ Lett.\ {\bf 83} (1999) 2498.

\bibitem{Madison00}
K.W.~Madison, F.~Chevy, W.~Wohlleben, and J.~Dalibard,
Vortex Formation in a Stirred Bose-Einstein Condensate,
Phys.\ Rev.\ Lett.\ {\bf 84} (2000) 806.

\bibitem{needcite} 
A. A. Abrikosov,
On the magnetic properties of superconductors of the second group,
Sov.\ Phys.\ JETP {\bf 5} (1957) 1174--1182.

\bibitem{vle} J. R. Abo-Shaeer, C. Raman, J. M. Vogels, W. Ketterle, 
Observation of vortex lattices in Bose-Einstein condensates, 
Science {\bf 292} (2001) 476--479.

\bibitem{needcite2} J. M. Kosterlitz and D. J. Thouless,
Ordering, metastability and phase
transitions in two-dimensional systems,
J. Phys. C: Solid State Physics {\bf 6} (1973) 1181--1203.

\bibitem{advanced} J. M. Kosterlitz, 
Topological defects and phase transitions
Rev. Mod. Phys. {\bf 89} (2017) 040501.  
%

\bibitem{dalikruger} Z. Hadzibabic, P. Kr{\"u}ger, M. Cheneau,
B. Battelier, and J. Dalibard, Berezinskii-Kosterlitz-Thouless crossover in a trapped
atomic gas, Nature {\bf 441} (2006) 1118--1121.


\bibitem{negretti} A. Negretti and C. Henkel,
Enhanced phase sensitivity and soliton formation in an integrated BEC interferometer, 	
J.\ Phys.\ B: At.\ Mol.\ Opt.\ Phys.\ {\bf 37} (2004) L385--L390.




\bibitem{jeff} J. Steinhauer, Observation of quantum Hawking radiation 
and its entanglement in an analogue black hole, Nature Phys. {\bf 12} (2016) 959–-965.

\bibitem{savage} T. R. Slatyer, C. M. Savage,
Superradiant scattering from a hydrodynamic vortex,
Class.\ Quantum Grav.\ {\bf 22} (2005) 3833-3839.

\bibitem{dasgupta} P. Das Gupta, E. Thareja,
On the possibility of generating Supermassive Black Holes from self-gravitating Bose-Einstein Condensates comprising of Bosonic Dark Matter,
Class.\ Quantum Grav.\ {\bf 34} (2017) 035006. 

\bibitem{gretchen} S. Eckel, A. Kumar, T. Jacobson, I.B. Spielman, G.K. Campbell,
A supersonically expanding Bose-Einstein condensate: an expanding universe in the lab,
Phys.\ Rev.\ X {\bf 8} (2018) 021021.




\bibitem{Kivshar-LutherDavies} 
Yu.S. Kivshar and B. Luther-Davies,
Dark optical solitons: physics and applications,
Phys.\ Rep.\ {\bf 298} (1998) 81--197.

\bibitem{djf} D. J. Frantzeskakis,
Dark solitons in atomic Bose-Einstein condensates: from theory to experiments,
J.\ Phys.\ A Math.\ Theor.\ {\bf 43} (2010) 213001.

\bibitem{Pismen1999} 
L. M. Pismen, 
Vortices in Nonlinear Fields (Clarendon, UK, 1999).

\bibitem{YSKPiO} 
Yu. S. Kivshar, J. Christou, V. Tikhonenko, B. Luther-Davies and L. Pismen,
Dynamics of optical vortex solitons,
Optics Comm.\ {\bf 152} (1998) 198--206.

\bibitem{fetter1} 
A. L. Fetter and A. A. Svidzinsky,
Vortices in a trapped dilute Bose-Einstein condensate,
J.\ Phys.: Condens.\ Matter {\bf 13} (2001) R135--R194.

\bibitem{fetter2} 
A. L. Fetter,
Rotating trapped Bose-Einstein condensates,
Reviews of Modern Physics {\bf 81} (2009) 647--691.

\bibitem{kuzne} E. A. Kuznetsov and S. K. Turitsyn,
Instability and collapse of solitons in media with a defocusing nonlinearity,
Zh.\ Eksp.\ Teor.\ Fiz.\ {\bf 94} (1988) 119--129 
[Sov.\ Phys.\ JETP {\bf 67} (1988) 1583--1588].

\bibitem{kidep} Yu. S. Kivshar and D.E. Pelinovsky,
Self-focusing and transverse instabilities of solitary waves,
Phys.\ Rep.\ {\bf 331} (2000) 117--195.

\bibitem{yskchristou} V. Tikhonenko, J. Christou, B. Luther-Davies, and Yu. S. Kivshar,
Observation of vortex solitons created by the instability of dark soliton stripes, 
Opt. Lett. {\bf 21} (1996) 1129--1131.

\bibitem{watching} B. P. Anderson, P. C. Haljan, C. A. Regal, D. L. Feder, 
L. A. Collins, C. W. Clark, and E. A. Cornell,
Watching Dark Solitons Decay into Vortex Rings in a Bose-Einstein Condensate,
Phys.\ Rev.\ Lett.\ {\bf 86} (2001) 2926--2929.


%

\bibitem{Williams99}
J. E.~Williams and M. J.~Holland,
Preparing topological states of a Bose-Einstein condensate,
Nature {\bf 401} (1999) 568.






\bibitem{kett99} 
R. Onofrio, C. Raman, J. M. Vogels, J. R. Abo-Shaeer, A. P. Chikkatur, 
and W. Ketterle,
Observation of Superfluid Flow in a Bose-Einstein Condensed Gas,
Phys.\ Rev.\ Lett.\ {\bf 85} (2000) 2228.

\bibitem{BPA3} 
T. W. Neely, E. C. Samson, A. S. Bradley, M. J. Davis, and B. P. Anderson,
Observation of Vortex Dipoles in an Oblate Bose-Einstein Condensate,
Phys.\ Rev.\ Lett.\ {\bf 104} (2010) 160401.


\bibitem{S2Ket1}
A. E. Leanhardt, A. G\"{o}rlitz, A. P. Chikkatur, D. Kielpinski, Y. Shin, D. E. Pritchard, 
and W. Ketterle, 
Imprinting Vortices in a Bose-Einstein Condensate using Topological Phases,
Phys.\ Rev.\ Lett.\ {\bf 89} (2002) 190403.


\bibitem{BPAPRL}
D. R.~Scherer, C. N.~Weiler, T. W.~Neely, and B. P.~Anderson,
Vortex Formation by Merging of Multiple Trapped Bose-Einstein Condensates,
Phys.\ Rev.\ Lett.\ {\bf 98} (2007) 110402.



\bibitem{BPA} 
C. N. Weiler, T. W. Neely, D. R. Scherer,
A. S. Bradley, M. J. Davis, and B. P. Anderson,
Spontaneous vortices in the formation of Bose-Einstein condensates,
Nature {\bf 455} (2008) 948--951.

\bibitem{dshall} 
D.V. Freilich, D.M. Bianchi, A.M. Kaufman, T.K. Langin, and D.S. Hall, 
Real-Time Dynamics of Single Vortex Lines and Vortex Dipoles in a Bose-Einstein Condensate,
Science {\bf 329} (2010) 1182--1185.


\bibitem{dshall1} 
S. Middelkamp, P.J. Torres, P.G. Kevrekidis, D.J. Frantzeskakis, R. Carretero-Gonz{\'a}lez, P. Schmelcher, D.V. Freilich, and D.S. Hall, 
Guiding-center dynamics of vortex dipoles in Bose-Einstein condensates,
Phys.\ Rev.\ A {\bf 84} (2011) 011605(R).

\bibitem{dshall2} 
P.J. Torres, P.G. Kevrekidis, D.J. Frantzeskakis, R. Carretero-Gonz{\'a}lez, 
P. Schmelcher, and D.S. Hall,
Dynamics of vortex dipoles in confined Bose-Einstein condensates,
Phys.\ Lett.\ A {\bf 375} (2011) 3044--3050.


\bibitem{dshall3} 
R. Navarro, R. Carretero-Gonz{\'a}lez, P. J. Torres, P. G. Kevrekidis, D. J. Frantzeskakis, 
M. W. Ray, E. Altunta\c{s}, and D. S. Hall,
Dynamics of a Few Corotating Vortices in Bose-Einstein Condensates
Phys.\ Rev.\ Lett.\ {\bf 110} (2013) 225301.

\bibitem{ex16_2} B. Gertjerenken, P. G. Kevrekidis, R. Carretero-Gonz{\'a}lez, and B. P. Anderson,
Generating and manipulating quantized vortices on-demand in a Bose-Einstein condensate: 
A numerical study, Phys.\ Rev.\ A {\bf 93} (2016) 023604.

\bibitem{ex16_3b} E.C. Samson, K.E. Wilson, Z.L. Newman, and B.P. Anderson,
Deterministic creation, pinning, and manipulation of quantized vortices in a Bose-Einstein condensate,
Phys.\ Rev.\ A {\bf 93} (2016) 023603.


\bibitem{nepom1} 
Yu.S. Kivshar, A. Nepomnyashchy, V. Tikhonenko, J. Christou, B. Luther-Davies,
Vortex-stripe soliton interactions, 
Opt.\ Lett.\ {\bf 25} (2000) 123--125.

\bibitem{nepom2} D. Neshev, A. Nepomnyashchy, Yu.S. Kivshar,
Nonlinear Aharonov-Bohm Scattering by Optical Vortices,
Phys.\ Rev.\ Lett.\ {\bf 87} (2001) 043901.

\bibitem{abeffect} Y. Aharonov and D. Bohm, 
Significance of electromagnetic potentials in quantum theory, 
Phys, Rev. {\bf 115} (1959) 485--491. 

\bibitem{nepom3} M. V. Berry, R. G. Chambers, M. D. Large, C. Upstill,
and J. C. Walmsley, Wavefront dislocations in the Aharonov-Bohm effect and its
water wave analogue, Eur. J. Phys. {\bf 1} (1980) 154--162. 

\bibitem{rdsus} G. Theocharis, D. J. Frantzeskakis, P. G. Kevrekidis, B. A. Malomed, 
and Yu. S. Kivshar, Ring dark solitons and vortex necklaces in Bose-Einstein condensates, 
Phys. Rev. Lett. {\bf 90} (2003) 120403.

\bibitem{kamchatnov} A. M. Kamchatnov and S. V. Korneev,
Dynamics of ring dark solitons in Bose-Einstein condensates
and nonlinear optics, Phys.\ Lett.\ A {\bf 374} (2010) 4625--4628.

\bibitem{smirnov} V. A. Mironov, A. I. Smirnov, and L. A. Smirnov, 
Dynamics of vortex structure formation during the evolution
of modulation instability of dark solitons, J. Exp. Theor. Phys. {\bf 112} (2011) 46-59.

\bibitem{siambook}
P. G. Kevrekidis, D. J. Frant\-zes\-ka\-kis, and R. Carretero-Gonz{\'a}lez,
{\sl The defocusing nonlinear Schr{\"o}dinger equation: from dark solitons
and vortices to vortex rings},
SIAM, Philadelphia (2015).



\bibitem{aipaper} P. G. Kevrekidis, Wenlong Wang,
  R. Carretero-Gonz{\'a}lez, and D. J. Frantzeskakis,
Adiabatic Invariant Approach to Transverse Instability: Landau Dynamics of Soliton Filaments,
Phys.\ Rev.\ Lett.\ {\bf 118} (2017) 244101.


\bibitem{rcg:105}
  Wenlong Wang, P.G. Kevrekidis, R. Carretero-Gonz{\'a}lez,
  D.J. Frantzeskakis, Tasso J. Kaper, and Manjun Ma,
Stabilization of ring dark solitons in Bose-Einstein condensates,
Phys.\ Rev.\ A {\bf 92} (2015) 033611.

\bibitem{babev} Wenlong Wang, P. G. Kevrekidis, and Egor Babaev,
  Ring dark solitons in three-dimensional Bose-Einstein condensates,
  Phys. Rev. A {\bf 100} (2019) 053621. 

\bibitem{george}
  P. G. Kevrekidis, Wenlong Wang, G. Theocharis, D. J. Frantzeskakis,
  R. Carretero-Gonz{\'a}lez, and B. P. Anderson,
  Dynamics of interacting dark soliton stripes,
  Phys. Rev. A \textbf{100} (2019) 033607.

\bibitem{bigelow} M. Jayaseelan, J. D. Murphree, J. T. Schultz,
A. Hansen, and N. P. Bigelow, Dark soliton rings in a Bose--Einstein condensate 
using Raman imprinting techniques,
    \url{https://meetings.aps.org/Meeting/DAMOP17/Event/300601}.

\bibitem{cct} C. Cohen-Tannoudji, B. Diu, and F. Lalo{\"e}, 
{\it Quantum Mechanics} (Wiley, New York, 2006).
  
\bibitem{DSS} 
W. Wang, P. G. Kevrekidis, R. Carretero-Gonz{\'a}lez, and D. J. Frantzeskakis,
Dark spherical shell solitons in three-dimensional Bose-Einstein condensates: Existence, stability, and dynamics,
Phys. Rev. A {\bf 93} (2016) 023630.

\bibitem{PW2}
R. Koll\'{a}r and R. L. Pego,
Spectral Stability of Vortices in Two-Dimensional Bose-Einstein Condensates via the Evans Function and Krein Signature,
Appl. Math. Res. eXpress {\bf 2012} (2011) 1.

\bibitem{gth} G. Theocharis, P. Schmelcher, M. K. Oberthaler, P. G. Kevrekidis, 
and D. J. Frantzeskakis, Lagrangian approach to the dynamics of dark matter-wave solitons,  
Phys. Rev. A {\bf 72} (2005) 023609. 



\bibitem{ourmarkus3}
G. Theocharis, A. Weller, J. P. Ronzheimer, C. Gross,
M. K. Oberthaler, P. G. Kevrekidis, and D.J. Frantzeskakis,
Multiple atomic dark solitons in cigar-shaped Bose-Einstein condensates,
Phys.\ Rev.\ A {\bf 81} (2010) 063604.

\bibitem{pubig} H. Pu, C.K. Law, J.H. Eberly, N.P. Bigelow,
Coherent disintegration and stability of vortices in trapped Bose
condensates,
Phys. Rev. A {\bf 59} (1999) 1533--1537.

\bibitem{carr}
G. Herring, L.D. Carr, R. Carretero-Gonz{\'a}lez,  P.G. Kevrekidis, and D.J. Frant\-zes\-ka\-kis,
Radially Symmetric Nonlinear States of Harmonically Trapped Bose-Einstein Condensates,
Phys.\ Rev.\ A {\bf 77} (2008) 023625.

\bibitem{rcg:99}
K.J.H. Law, T.W. Neely, P.G. Kevrekidis, B.P. Anderson, A.S. Bradley,
and
R. Carretero-Gonz{\'a}lez,
Dynamic and Energetic Stabilization of Persistent Currents in Bose-Einstein Condensates,
Phys.\ Rev.\ A {\bf 89} (2014) 053606.

\bibitem{bisset}
  R. N. Bisset, Wenlong Wang, C. Ticknor, R. Carretero-Gonz{\'a}lez,
  D. J. Frantzeskakis, L. A. Collins, and P. G. Kevrekidis,
  Robust vortex lines, vortex rings, and hopfions in three-dimensional
  Bose-Einstein condensates,
  Phys. Rev. A {\bf 92} (2015), 063611.

\bibitem{danaila04} L.-C. Crasovan, V. M. P{\'e}rez-Garc{\i}a,
  I. Danaila, D. Mihalache, and L. Torner,
Three-dimensional parallel vortex rings in Bose-Einstein condensates,
  Phys. Rev. A {\bf 70} (2004) 033605.

%
%





\end{thebibliography}
\end{document}